\documentclass[10pt]{article} 
\usepackage{amssymb}
\usepackage{amsmath}
\usepackage{amsthm}
\usepackage{latexsym} 
\usepackage[dvips]{epsfig}

\theoremstyle{plain}
 
\newtheorem{lemma}{Lemma}
\newtheorem{theorem}{Theorem}
\newtheorem*{assumptions}{Assumptions}
\newtheorem*{conjecture}{Conjecture}
\newtheorem{corollary}{Corollary}
\newtheorem*{main}{Theorem}

\def\O{\mathcal{O}}

\font\SYM=msbm10 
\newcommand{\Real}{\mbox{\SYM R}}
\newcommand{\Complex}{\mbox{\SYM C}}

\font\tenscr=rsfs10 scaled1100
\font\sevenscr=rsfs7 
\font\fivescr=rsfs5 
\skewchar\tenscr='177 
\skewchar\sevenscr='177
\skewchar\fivescr='177 
\newfam\scrfam 
\textfont\scrfam=\tenscr
\scriptfont\scrfam=\sevenscr 
\scriptscriptfont\scrfam=\fivescr

\def\scri{{\fam\scrfam I}}

\begin{document}



\title{A new class of obstructions to the smoothness of null infinity}

\author{Juan Antonio Valiente Kroon \thanks{E-mail address:
 {\tt jav@aei-potsdam.mpg.de}} \\
 Max Planck Institut f\"ur Gravitationphysik,\\ Albert Einstein Institut,\\
Am M\"{u}hlemberg 1, 14476 Golm,\\ Germany.}

\maketitle

\begin{abstract}
  Expansions of the gravitational field arising from the development
  of asymptotically Euclidean, time symmetric, conformally flat
  initial data are calculated in a neighbourhood of spatial and null
  infinities up to order 6. To this ends a certain representation of
  spatial infinity as a cylinder is used. This set up is based on the
  properties of conformal geodesics. It is found that these expansions
  suggest that null infinity has to be non-smooth unless the
  Newman-Penrose constants of the spacetime, and some other higher
  order quantities of the spacetime vanish. As a consequence of these
  results it is conjectured that similar conditions occur if one were
  to take the expansions to even higher orders. Furthermore, the
  smoothness conditions obtained suggest that if a time symmetric
  initial data which is conformally flat in a neighbourhood of spatial
  infinity yields a smooth null infinity, then the initial data must
  in fact be Schwarzschildean around spatial infinity.
\end{abstract}

\section{Introduction}
  
Penrose introduced the seminal idea that the gravitational field of
isolated systems can be conveniently described by means of the notion
of asymptotic simplicity \cite{Pen63} \footnote{It will be assumed
  that the reader is familiar with the ideas of the so-called
  conformal framework to describe the properties of isolated bodies
  and the concept of asymptotic flatness. For a recent review, the
  reader is remitted to \cite{Fri02a}.}. Central to the concept of
asymptotic simplicity is the idea ---expectation--- that the conformal
boundary of the spacetime ---null infinity, $\scri$--- should possess
a smooth differentiable structure. This approach to the description of
isolated bodies in General Relativity is usually known as
\emph{Penrose's proposal} ---see e.g. \cite{Fri99,Fri02a}.  Static and
stationary spacetimes have been shown to be (weakly) asymptotically
simple, with a smooth null infinity \cite{Dai01b}. However, the main
purpose behind introducing the concept of asymptotic simplicity is to
provide a suitable framework for the discussion of radiation. In spite
of its elegance and aesthetical appeal, Penrose's proposal is of
little use if one is not able to prove that there exists a big family
of non-trivial (in the radiative sense) asymptotically simple
solutions to the Einstein Field Equations. A programme aimed to
investigate the existence of such solutions, and provide a conclusive
``answer'' to Penrose's proposal has been started by Friedrich ---see
e.g. \cite{Fri81,Fri81a,Fri83,Fri86a,Fri86b,Fri88,Fri98a}. His
strategy is based on the use of the so-called Conformal Field
Equations, which allow us to work and prove existence statements
directly in the conformally rescaled, ``unphysical'' spacetime. Along
this lines, Friedrich has been able to prove a semiglobal existence
result that ensures that hyperboloidal initial data close to Minkowski
data yields an asymptotically simple development which includes the
point $i^+$ (future timelike infinity) \cite{Fri86b}. Recently,
Chru\'{s}ciel \& Delay \cite{ChrDel02} ---retaking an old idea by
Cutler \& Wald \cite{CutWal88}--- using a refined version of some
initial data set constructed by Corvino \cite{Cor00} have been able to
prove the existence of a big class of non-trivial (radiative)
asymptotically simple spacetimes.  Corvino's initial data is
constructed so that it is Schwarzschildean in a neighbourhood of
spatial infinity, $i^0$. This means that the radiation content in the
spacetime arising from the development of the data is somehow special.
This can be seen directly from the fact that the Newman-Penrose
constants of the spacetime are zero \cite{DaiVal02,Val02b}.
Chru\'{s}ciel \& Delay's result is no doubt very important. However,
it is not as general as one would like.
 
It has been suspected for a long time now that the region of spacetime
where null infinity and spatial infinity meet is somehow problematic
---see e.g.  \cite{Pen65a}. From the analysis of the hyperboloidal
initial value problem it turns out that the smoothness of null
infinity is preserved by the evolution if smooth data sufficiently
close to Minkowski data are prescribed. The latter indicates that
somehow the ``decision'' of having a smooth structure at null infinity
is made in an arbitrarily small neighbourhood of spatial infinity.  In
some sense, Corvino's data avoids all the intricacies and
complications of this region of spacetime by setting the asymptotic
end in the simplest way which is consistent with the presence of a
non-vanishing ADM mass on the initial hypersurface. In connection with
this, Friedrich \cite{Fri98a} has performed a detailed first analysis
of the behaviour of the gravitational field arising from
asymptotically Euclidean, time symmetric initial data in the region
where null infinity ``touches'' spatial infinity. By means of a novel
representation of spatial infinity in which the point $i^0$ of the
standard conformal picture is blown up to a cylinder $I$ --- the
cylinder at spatial infinity--- a certain regularity condition on the
Cotton-Bach tensor and symmetrised higher order derivatives of it has
been obtained. The hope was that this regularity condition would
ensure the smoothness of null infinity, at least in the region close
to $i^0$ and $\scri$. A subsequent analysis by Friedrich \&
K\'{a}nn\'{a}r \cite{FriKan00} of the first orders of some expansions
that can be obtained by evaluating the Conformal Field Equations at
$I$, lead to conjecture that Friedrich's regularity condition is the
only condition one has to impose on time symmetric, initial data
possessing an analytic compactification in order to obtain a
development with smooth null infinity \cite{Fri02a}. More precisely,

\begin{conjecture}[Friedrich, 2002]
There exists an integer $k_*>0$ such hat for given $k\leq k_*$ the time
evolution of an asymptotically Euclidean, time symmetric, conformally
smooth initial data set admits a conformal extension to null infinity
of class $C^k$ near spacelike infinity, if the Cotton-Bach
spinor\footnote{ The Cotton-Bach tensor $B_{ijk}$ is related to the
  Cotton-Bach spinor via:
\[
B_{ijk}\mapsto b_{abce} \epsilon_{df}+ b_{abdf}\epsilon_{ce}.
\]
This correspondence is carried out by the Infeld-van der Waerden symbols.}
 satisfies the condition,
\[
D_{(a_sb_s}\cdots D_{a_1b_1} b_{abcd)}(i)=0, \quad s=0,1,\ldots
\]
for a certain integer $s_*=s_*(k)$. If the extension is of class $C^\infty$
then the condition should hold to all orders.
\end{conjecture}

The objective of this paper is to provide a further insight into
the conjecture above. It will turn out that the
conjecture, as it stands, is false. In order to see why this is the
case, the expansions of K\'{a}nn\'{a}r \& Friedrich will be carried to
an even higher order. This requires the
implementation of the Conformal Field Equations on a computer algebra
system ({\tt Maple V}). It should be emphasized that despite the use
of the computer to perform the expansions, the results
here presented are exact up to the order carried
out. In order to simplify our discussion, the analysis will be
restricted to developments of time symmetric conformally flat initial
data. The time symmetry requirement stems from the fact that
Friedrich's analysis has only yet been carried for this
class of data. A similar analysis of initial data with non-vanishing
second fundamental form lies still in the future. However, some first steps
have already been carried out \cite{DaiFri01}. The
conformal flatness of the data ensures that the initial data satisfies
the regularity condition trivially. Again, the construction of
non-conformally flat data satisfying Friedrich's regularity condition
is a non-trivial endeavour whose undertaking will be left for future
studies. In the light of the results here presented, it turns out that
the restriction to the class of conformally flat data is not a
drawback. Furthermore, it is not hard to guess how the results could
generalise in the case of general time symmetric data.

The principal result of our investigation is the following,

\begin{main}[Main theorem]
  Necessary conditions for the development of initial data which are
  time symmetric, conformally flat in a neighbourhood $B_a(i)$ of
  (spatial) infinity to be smooth at the intersection of null infinity
  and spatial infinity are that the Newman-Penrose constants
  $G^{(5)}_k$, $k=0,\ldots,4$ and the higher order Newman-Penrose
  constants, $G^{(6)}_k$, $k=0,\ldots,6$ vanish.
\end{main}

A more precise formulation of the theorem, including the definition of the
Newman-Penrose constants and the higher order Newman-Penrose constants in
terms of the initial data will be given in the main text.

We note that the previous conditions are fulfilled by the
Schwarzschild initial data. The theorem has been obtained from the
analysis, up to order $p=6$ of expansions constructed from the
solutions of the transport equations induced by the conformal field
equations upon evaluation on the cylinder at spatial infinity. From
the evidence provided by the equations it is not unreasonable to
conjecture that similar conditions arise if one were to obtain
expansions to even higher orders. In \cite{Fri98a} it was shown that
the whole set up of the cylinder at spatial infinity is completely
regular for Schwarzschildean data. Thus, the hypothetic new conditions
must be satisfied by the Schwarzschild initial data.  It is on this
ground that the following conjecture is put forward:

\begin{conjecture}[New conjecture]
If an initial data set which is time symmetric and conformally flat in
a neighbourhood $B_a(i)$ of the point $i$ yields a development with a smooth
null infinity, then the initial data are in fact Schwarzschildean in
$B_a(i)$.
\end{conjecture}

Again, a more technical version of the conjecture is given in the main
text.
 
\bigskip The article is structured as follows: in section 2 a brief
summary of the description of spacetime in the neighbourhood of
spatial and null infinities in terms of the cylinder at spatial
infinity is given. This digest has the intention of providing all the
tools needed for the calculations described in the present article.
Particular attention is paid to the spatial 2-spinor formalism and to
the expansions of functions on $S^3$ in terms of unitary
representations of $SU(2,\Complex)$. The reader is, in any case,
remitted to \cite{Fri98a} for a more extensive discussion. In section
3, the conformal field equations written in the conformal geodesic
gauge are discussed. The initial data for the latter in the case of an
asymptotically Euclidean, time symmetric, conformally flat initial
hypersurface are described. The transport equations implied by the
conformal field equations on the cylinder at spatial infinity are
introduced in section 4. In order to properly motivate the
calculations carried out in later sections, a careful discussion of
their properties is undertaken. Section 5 contains the new results to
be presented in the article. Here a description of the solution of the
transport equations is done. Due to the large size of the expressions
involved, the description will be focused on what we believe are the
most relevant features of the solutions. However, it should be
emphasized that everything has been explicitly calculated. The main
conclusions extracted from the calculations is presented as our main
theorem. In order to understand the meaning of the conditions
presented in the main theorem, the Schwarzschild solution is discussed
in this context, and a conjecture is formulated. Finally, in section
6, some conclusions and extensions of the present work are considered.
There is also an appendix in which the implementation of the transport
equations and how to solve them in the computer algebra system {\tt
  Maple V} is briefly considered.

\section{Spacetime in a neighbourhood of spatial infinity}

Let $(\widetilde{M}, \widetilde{g}_{\mu\nu})$ be a vacuum spacetime
arising as the development of asymptotically Euclidean, time
symmetric, conformally analytic initial data
$(\widetilde{S},\widetilde{h}_{\alpha,\beta})$. Later, we will
restrict the class of initial data sets under discussion to those
conformally flat around infinity. Assume for simplicity that
$\widetilde{S}$ possesses only one asymptotic end. Let $i$ be the
infinity corresponding to that end.  The point $i$ is obtained by
conformally compactifying the initial hypersurface $\widetilde{S}$,
with an analytic conformal factor $\Omega$ which can be obtained from
solving the time symmetric constraint equations. The compact 3
dimensional manifold obtained in this way will be denoted by $S$, and
the conformally rescaled 3-metric by $h_{\alpha\beta}$.  Assume that
the 3-metric $h_{\alpha\beta}$ is analytic in an open ball $B_a(i)$ of
radius $a$ centered on $i$. Let $\rho$ denote the geodesic distance
along geodesics starting at $i$. The radius $a$ of the ball $B_a(i)$
is chosen such that $B_a(i)$ is geodesically convex.

Now, let $\widetilde{N}\cap\widetilde{M}$ be the domain of influence of the
ball $B_a(i)\cap S$. Intuitively, one expects $N$ to cover a region of
spacetime ``close to null and spatial infinities''. In reference \cite{Fri98a}
it has been shown that once the time symmetric constraint equations have been
solved, a certain gauge based on the properties of conformal geodesics can be
introduced. Let $\tau$ be the parameter along these curves. This gauge has the
property of producing a conformal factor $\Theta$ which can be in turn used to
rescale the region $\widetilde{N}$ to obtain a ``finite representation'' $N$,
of spacetime in a neighbourhood of spatial and null infinities. The relevant
conformal factor is given by,
\begin{equation}
\label{conformal_factor}
\Theta=\kappa^{-1} \Omega\left(1-\tau^2\frac{\kappa^2}{\omega^2}\right),
\end{equation}
where $\omega$ is given by,
\begin{equation}
\label{omega}
\omega=\frac{2\Omega}{\sqrt{|D_a\Omega D^{a}\Omega|}},
\end{equation}
and $\kappa$ is a smooth function such that $\kappa=\kappa'\rho$ with
$\kappa'(i)=1$. It contains the remaining piece of conformal freedom in our
setting.

Throughout this work, space 2-spinors will be systematically used. In
order to avoid problems with vanishing frame vectors on surfaces diffeomorphic
to spheres, our discussion will be carried not on $N$ but on a subbundle
$C_{a,\kappa}$ of the frame bundle over $N$. The subbundle $C_{a,\kappa}$ can
be shown to be a 5-dimensional submanifold of $\Real\times\Real \times SU(2,\Complex)$
with structure group $U(1)$. More precisely, we define $C_{a,\kappa}$ to be
given by,
\begin{equation}
C_{a,\kappa}=\left\{(\tau,\rho,t)\in \Real\times\Real \times SU(2,\Complex) |\; 0\leq
  \rho <a,\;\;-\frac{\omega}{\kappa}\leq \tau\leq
  \frac{\omega}{\kappa} \right\}. 
\end{equation}
The projection $\pi'$ of $C_{a,\kappa}$ into $N$ corresponds to the Hopf map
$SU(2,\Complex) \rightarrow SU(2,\Complex)/U(1)\approx S^2$. Scalar fields and tensorial fields
on $N$ are lifted to $C_{a,\kappa}$. Their ``angular'' dependence will be then
given in terms of functions of $t\in SU(2,\Complex)$.

The manifold $C_{a,\kappa}$ has the following important submanifolds,
\begin{subequations}
\begin{eqnarray}
&& I=\{(\tau,\rho,t)\in C_{a,\kappa} \; |\; \rho=0, \;\;|\tau|<1 \}, \\
&& I^\pm=\{(\tau,\rho,t)\in C_{a,\kappa} \;  |\;\rho=0, \;\;\tau=\pm 1\}, \label{degenerate_sets}\\
&& \scri^\pm=\left\{(\tau,\rho,t)\in C_{a,\kappa} \; |\; \rho>0 ,\;\;
\tau=\pm\frac{\omega}{\kappa}\right\}.
\end{eqnarray}   
\end{subequations}

\subsection{Space 2-spinors}
Consider the antisymmetric spinors $\epsilon_{ab}$, $\epsilon^{ab}$,
$a,b=0,1$. These satisfy $\epsilon_{01}=1$, $\epsilon^{01}=1$. Let
$\tau^{aa'}$ denote the tangent vector to the conformal geodesics parametrised
by $\tau$. We set,
\begin{equation}
\tau^{aa'}=\epsilon_{0}^{\phantom{0}a}\overline{\epsilon}_{0'}^{\phantom{0'}a'}+\epsilon_{1}^{\phantom{1}a}\overline{\epsilon}_{1'}^{\phantom{1'}a'}.
\end{equation}
In order to define introduce differential operators on $SU(2,\Complex)$ we consider the
basis
\begin{equation}
u_1=\frac{1}{2}\left(
\begin{array}{cc}
0 & i \\
i & 0 \\
\end{array}
\right), 
\quad
u_2=\frac{1}{2}\left(
\begin{array}{cc}
0 & -1 \\
1 & 0 \\
\end{array}
\right),
\quad
u_3=\frac{1}{2}\left(
\begin{array}{cc}
i & 0 \\
0 & -i \\
\end{array}
\right),
\end{equation}
of the Lie algebra of SU(2,\Complex). Note that in particular, $u_3$ generates
$U(1)$. Denote by $Z_i$, $i=1,2,3$ the real left invariant vector field
generated by $u_i$ on SU(2,\Complex). We define the, following complex vector fields
\begin{equation}
X_+=-(Z_2+iZ_1), \quad X_-=-(Z_2-iZ_1), \quad X=-2iZ_3.
\end{equation}
These are such that $[X_+,X_-]=-X$. With the help of $X_\pm$ one can construct
the following (frame) vectors fields,
\begin{equation}
c_{ab}=c^0_{aa'}\partial_\tau+c^1_{aa'}\partial_\rho+c^+_{aa'}X_+
+c^-_{aa'}X_-,
\end{equation}
on $C_{a,\kappa}$. The use of a space spinor formalism based on the vector
field $\sqrt{2}\partial_\tau=\tau^{aa'}c_{aa'}$ allows to perform our whole 
discussion in terms of quantities without primed indices. Accordingly, we
write
\begin{equation}
c_{aa'}=\frac{1}{\sqrt{2}}\tau_{aa'}\partial_\tau -
\tau^b_{\phantom{b}a'}c_{ab}\end{equation}
with
$c_{ab}=\tau_{(a}^{\phantom{(a}b'}c_{b)b'}=c^0_{ab}\partial_\tau+c^1_{ab}\partial_\rho+c^+_{ab}X_+
+c^-_{ab}X_-$. The connection is represented by coefficients $\Gamma_{abcd}$
which can be decomposed in the form,
\begin{equation}
\Gamma_{abcd}=\frac{1}{\sqrt{2}}\left( \xi_{abcd}
  -\chi_{(ab)cd}\right)-\frac{1}{2}\epsilon_{ab}f{cd},
\end{equation}
where the fields entering in the decomposition possess the following
symmetries: $\chi_{abcd}=\chi_{ab(cd)}$, $\xi_{abcd}=\xi_{(ab)(cd)}$,
$f_{ab}=f_{(ab)}$. The curvature will be described by the rescaled conformal
Weyl spinor $\phi_{abcd}=\phi_{(abcd)}$, and by the spinor field
$\Theta_{abcd}=\Theta_{ab(cd)}$ which encodes information relative to the Ricci
part of the curvature. For the purpose of writing the field equations it
will be customary to consider it decomposed in terms of
$\Theta_{g\phantom{g}cd}^{\phantom{g}g}$ and $\Theta_{(ab)cd}$.

For latter use it is noted that an arbitrary four indices spinor $X_{abcd}$
can be written in terms of the ``elementary spinors'' $\epsilon^i_{abcd}$,
$\epsilon_{ac}x_{bd}+\epsilon_{bd}x_{ac}$,
$\epsilon_{ac}y_{bd}+\epsilon_{bd}y_{ac}$,
$\epsilon_{ac}z_{bd}+\epsilon_{bd}z_{ac}$, and $h_{abcd}$ where,
\begin{equation}
x_{ab}=\sqrt{2}\epsilon_{(a}^{\phantom{a}0}\epsilon_{b)}^{\phantom{b)}1},
\quad y_{ab}=-\frac{1}{\sqrt{2}} \epsilon_{a}^{\phantom{a}1}
\epsilon_{b}^{\phantom{b}1}, \quad   z_{ab}=\frac{1}{\sqrt{2}} \epsilon_{a}^{\phantom{a}0}
\epsilon_{b}^{\phantom{b}0},
\end{equation}
and,
\begin{equation}
\epsilon^i_{abcd}=\epsilon_{(a}^{\phantom{(a}(e}\epsilon_b^{\phantom{b}f}\epsilon_c^{\phantom{c}g}\epsilon_{d)}^{\phantom{d)}h)_i},
\quad h_{abcd}=-\epsilon_{a(c}\epsilon_{d)b}.
\end{equation}  
The notation $(abcd)_i$ means that the indices are to be symmetrised and then
$i$ of them set to $1$. We write,
\begin{eqnarray}
&& X_{abcd}= X_0 \epsilon^0_{abcd}+X_1 \epsilon^1_{abcd}+X_2
\epsilon^2_{abcd}+X_3 \epsilon^3_{abcd}+X_4 \epsilon^4_{abcd} \nonumber \\
&& \phantom{ X_{abcd}=XX} +X_x (\epsilon_{ac}x_{bd}+\epsilon_{bd}x_{ac})+
X_y(\epsilon_{ac}y_{bd}+\epsilon_{bd}y_{ac}) +
X_z(\epsilon_{ac}z_{bd}+\epsilon_{bd}z_{ac}) \nonumber \\
&&\phantom{ X_{abcd}=XXXX}+ X_h h_{abcd}.
\end{eqnarray}

\subsection{Expansions of functions on $C_{a,\kappa}$}

In order to obtain our expansions of the gravitational field around
spatial infinity, we will require to decompose the functions arising
into their diverse spherical (harmonic) sectors. Any function real
analytic complex value function $f$ on $SU(2,\Complex)$ can be
expanded in terms of some functions $\sqrt{m+1} \;
T_{m\phantom{k}j}^{\phantom{m}k}$ forming a complete set in
$L^2(\mu,SU(2,\Complex))$ where $\mu$ is the standard Haar measure in
$SU(2,\Complex)$. One has,
\[
f(t)=\sum^\infty_{m=0}\sum^m_{j=0}\sum^m_{k=0}f_{m,k,j} T_{m\phantom{k}j}^{\phantom{m}k}.
\]
The functions $T_{m\phantom{k}j}^{\phantom{m}k}$ can be shown to be
related with the standard spin-weighted spherical harmonics. Using the
fact that the group $SU(2,\Complex)$ is diffeomorphic to $S^3$, one
can use the coordinates $(\zeta,\overline{\zeta},\alpha)$ to write a
given $t\in SU(2,\Complex)$ as,
\begin{equation}
t \in \frac{1}{\sqrt{1+\zeta\overline{\zeta}}}
\left(
\begin{array}{cc}
e^{i\alpha} & ie^{-i\alpha}\zeta \\
ie^{i\alpha}\overline{\zeta} & e^{i\alpha} \\
\end{array}
\right).
\end{equation} 
The latter yields the following correspondence rule between the
functions $T_{m\phantom{k}j}^{\phantom{m}k}$ and the spin-weighted
spherical harmonics ${}_sY_{ln}$:
\begin{equation}
\label{Y_to_T}
{}_sY_{nm} \mapsto (-i)^{s+2n-m}\sqrt{\frac{2n+1}{4\pi}}T_{2n\phantom{n-m}n-s}^{\phantom{2n}n-m}.
\end{equation}

The operators $X_\pm$ and $X$ (related to the NP $\eth$ and
$\overline{\eth}$ operators) introduced in the
previous sections can be seen to yield, upon application to the  $
T_{m\phantom{k}j}^{\phantom{m}k}$ functions the following,
\begin{subequations}
\begin{eqnarray}
&&X_+ T_{m\phantom{k}j}^{\phantom{m}k}=\beta_{m,j}  T_{m\phantom{k}j-1}^{\phantom{m}k}, \\ 
&&X_- T_{m\phantom{k}j}^{\phantom{m}k}=-\beta_{m,j+1}
T_{m\phantom{k}j+1}^{\phantom{m}k}, \qquad \mbox{with }
\beta_{m,j}=\sqrt{j(m-j+1)}.
\end{eqnarray}
\end{subequations}
A function $f$ on $SU(2,\Complex)$
is said to have spin weight $s$ if $Xf=2sf$. This definition can be
readily extended to functions on $C_{a,\kappa}$. As it will be seen later, all
the quantities we will work with will have a well defined spin weight.
Let $f$ be an analytic function $f$ on $C_{a,\kappa}$ with an integer spin
weight $s$.

Now, consider a spinorial symmetric analytic function on $C_{a,\kappa}$,
$\lambda_{a_1\cdots a_{2r}}$ with essential components
$\lambda_j=\lambda_{(a_1,\cdots\,a_{2r})_j}$, $0\leq j \leq 2r$, of spin weight
$s=r-j$. Then, the components of the function will possess expansions of the
form,
\begin{equation}
\lambda_j=\sum_{p=0}^\infty \lambda_{j,p}\rho^p,
\end{equation}
where the coefficients $\lambda_{j,p}$ can in turn be decomposed in
terms of the functions $T_{m\phantom{k}j}^{\phantom{m}k}$, as
\begin{equation}
\lambda_{j,p}=\sum_{q=|r-j|}^{q(p)}\sum_{k=0}^{2q}
\lambda_{j,p;2q,k}T_{2q\phantom{k}q-r+j}^{\phantom{2q}k},
\end{equation}
where $0\leq |r-j| \leq q(p) \leq \infty$. An expansion of the latter
type will be referred to as \emph{an expansion of type} $q(p)$. 

The conformal field equations are nonlinear. Thus, when expanding
them, one finds products of $T$-functions. These products can in turn
be reexpressed as a linear combination of $T$'s. More precisely, one has:

\begin{lemma}
\textbf{Multiplying $T$ functions.} The following holds,
\begin{eqnarray}
&&T_{2n_1\phantom{k_1}l_1}^{\phantom{2n_1}k_1} \times
T_{2n_2\phantom{k_2}l_2}^{\phantom{2n_2}k_2} = \nonumber \\
&&\phantom{XXX}\sum_{n=q_0}^{n_1+n_2}(-1)^{n+n_1+n_2}
 C(n_1,n_1-l_1;n_2,n_2-l_2;n,n_1+n_2-l_1-l_2) \nonumber \\
&&\phantom{XXXXXXX}\times C(n_1,n_1-k_1;n_2,n_2-k_2;n,n_1+n_2-k_1-k_2)
\nonumber\\
&&\phantom{XXXXXXXXXXX}\times
T_{2n\phantom{n+k_1+k_2-n_1-n_2}n+l_1+l_2-n_1-n_2}^{\phantom{2n}n+k_1+k_2-n_1-n_2}, \label{monster}
\end{eqnarray}
where $q_0=max\{|n_1-n_2|, n_1+n_2-k_1-k_2,n_1+n_2-l_1-l_2\}$, and
$C(l_1,m_1;l_2,m_2;l,m)$ denotes the standard Clebsch-Gordan
coefficients of $SU(2,\Complex)$\,\,\,\footnote{Some other used notations in the physics
  literature are:
\begin{eqnarray*}
&&C(l_1,m_1;l_2,m_2;l,m)\equiv <l_1,l_2;m_1,m_2|l,m>, \nonumber \\
&&\phantom{C(l_1,m_1;l_2,m_2;l,m)}\equiv C(l_1,l_2,l|m_1,m_2,m).
\end{eqnarray*}
.}
\end{lemma}

\section{The conformal evolution equations}

Using the conformal geodesic gauge and the two spinor decomposition, it can be
shown that the extended conformal field equations given in \cite{Fri98a} imply
the following evolution equations for the unknowns
$v=\left(c^\mu_{ab},\xi_{abcd},f_{ab},\chi_{(ab)cd},\Theta_{(ab)cd},\Theta_{g\phantom{g}cd}^{\phantom{g}g}
\right)$, $\mu=0,1,\pm$, 
\begin{subequations}
\begin{eqnarray}
&&\partial_\tau c^0_{ab}=-\chi_{(ab)}^{\;\;\;\;\;\;ef}c^{0}_{ef}-f_{ab}, \label{p1} \\
&&\partial_\tau c^\alpha_{ab}=-\chi_{(ab)}^{\;\;\;\;\;\;ef}c^\alpha_{ef}, \label{p2}\\
&&\partial_\tau \xi_{abcd}=-\chi_{(ab)}^{\;\;\;\;\;\;ef}\xi_{efcd}+\frac{1}{\sqrt{2}}(\epsilon_{ac}\chi_{(bd)ef}+\epsilon_{bd}\chi_{(ac)ef})f^{ef} \nonumber\\
&&\;\;\;\;\;\;\;\;\;\;\;\;\;\;\;\; -\sqrt{2}\chi_{(ab)(c}^{\phantom{(ab)(c}e}f_{d)e}-\frac{1}{2}(\epsilon_{ac}\Theta_{f\;\;bd}^{\;\;f}+\epsilon_{bd}\Theta_{f\;\;ac}^{\;\;f})-i\Theta\mu_{abcd}, \label{p3} \\
&&\partial_\tau f_{ab}=-\chi_{(ab)}^{\;\;\;\;\;\;ef}f_{ef}+\frac{1}{\sqrt{2}}\Theta_{f\;\;ab}^{\;\;f}, \label{p4} \\
&&\partial_\tau \chi_{(ab)cd}=-\chi_{(ab)}^{\;\;\;\;\;\;ef}\chi_{efcd}-\Theta_{(cd)ab}+\Theta\eta_{abcd}, \label{p5} \\
&&\partial_\tau\Theta_{(ab)cd}=-\chi_{(ab)}^{\;\;\;\;\;\;ef}\Theta_{(ab)ef}-\partial_\tau\Theta\eta_{abcd}+i\sqrt{2}d^e_{\;(a}\mu_{b)cde}, \label{p6} \\
&&\partial_\tau\Theta_{g\;\;ab}^{\;\;g}=-\chi_{(ab)}^{\;\;\;\;\;\;ef}\Theta_{g\phantom{g}ef}^{\phantom{g}g}+\sqrt{2}d^{ef}\eta_{abef}, \label{p7}
\end{eqnarray}
\end{subequations}
where 
\begin{equation}
\eta_{abcd}=\frac{1}{2}(\phi_{abcd}+\tau_a^{\phantom{a}a'}\tau_b^{\phantom{b}b'}\tau_c^{\phantom{c}c'}\tau_d^{\phantom{d}d'}\overline{\phi}_{a'b'c'd'}), \quad
\mu_{abcd}=-\frac{i}{2}(\phi_{abcd}-\tau_a^{\phantom{a}a'}\tau_b^{\phantom{b}b'}\tau_c^{\phantom{c}c'}\tau_d^{\phantom{d}d'}\overline{\phi}_{a'b'c'd'}),
\end{equation}
denoting respectively the electric and magnetic part of of $\phi_{abcd}$. The
quantities $\Theta$, $\partial_\tau\Theta$ and $d_{ab}$, given by formulae
(\ref{conformal_factor}) and (\ref{one_form})  are known directly
from the initial data. Thus, the equations (\ref{p1})-(\ref{p7}) are
essentially ordinary differential equations for the components of the vector
$v$. 

The most important part of the propagation equations corresponds to 
the evolution equations for the spinor $\phi_{abcd}$ derived from the
Bianchi identities \emph{Bianchi propagation equations}:
\begin{subequations} 
\begin{eqnarray}
&&(\sqrt{2}-2c^0_{01})\partial_\tau\phi_0+2c^0_{00}\partial_\tau\phi_1-2c^\alpha_{01}\partial_{\alpha}\phi_0+2c^\alpha_{00}\partial_\alpha\phi_1 \nonumber \\
&&\;\;\;= (2\Gamma_{0011}-8\Gamma_{1010})\phi_0+(4\Gamma_{0001}+8\Gamma_{1000})\phi_1-6\Gamma_{0000}\phi_2, \label{b0}\\
&&\sqrt{2}\partial_\tau\phi_1-c^0_{11}\partial_\tau\phi_0+c^0_{00}\partial_\tau\phi_2-c^\alpha_{11}\partial_{\alpha}\phi_0+c^\alpha_{00}\partial_{\alpha}\phi_2\nonumber\\
&&\;\;\;=-(4\Gamma_{1110}+f_{11})\phi_0+(2\Gamma_{0011}+4\Gamma_{1100}-2f_{01})\phi_1
+3f_{00}\phi_2-2\Gamma_{0000}\phi_3, \label{b1} \\
&&\sqrt{2}\partial_\tau\phi_2-c^0_{11}\partial_\tau\phi_1+c^0_{00}\partial_\tau\phi_3-c^\alpha_{11}\partial_{\alpha}\phi_1+c^\alpha_{00}\partial_{\alpha}\phi_3\nonumber \\
&&\;\;\;=-\Gamma_{1111}\phi_0-2(\Gamma_{1101}+f_{11})\phi_1+3(\Gamma_{0011}+\Gamma_{1100})\phi_2-2(\Gamma_{0001}-f_{00})\phi_3-\Gamma_{0000}\phi_4, \label{b2}\\
&&\sqrt{2}\partial_\tau\phi_3-c^0_{11}\partial_\tau\phi_2+c^0_{00}\partial_\tau\phi_4-c^\alpha_{11}\partial_{\alpha}\phi_2+c^\alpha_{00}\partial_{\alpha}\phi_4\nonumber \\
&&\;\;\;=-2\Gamma_{1111}\phi_1
-3f_{11}\phi_2+(2\Gamma_{1100}+4\Gamma_{0011}+2f_{01})\phi_3-(4\Gamma_{0001}-f_{00})\phi_4,
\label{b3}\\
&&(\sqrt{2}+2c^0_{01})\partial_\tau\phi_4-2c^0_{11}\partial_\tau\phi_3+2c^\alpha_{01}\partial_\alpha\phi_4-2c^\alpha_{11}\partial_\alpha\phi_3 \nonumber \\
&&\;\;\;=-6\Gamma_{1111}\phi_2+(4\Gamma_{1110}+8\Gamma_{0111})\phi_3
+(2\Gamma_{1100}-8\Gamma_{0101})\phi_4. \label{b4}
\end{eqnarray}
\end{subequations}
To the latter we add a set of three equations, also implied by the Bianchi
identities which we refer as to the \emph{Bianchi constraint
  equations,}
\begin{subequations}  
\begin{eqnarray}
&&c^0_{11}\partial_\tau\phi_0-2c^0_{01}\partial_\tau\phi_1+c^0_{00}\partial_\tau \phi_2 + c^\alpha_{11}\partial_\alpha\phi_0-2c^\alpha_{01}\partial_\alpha\phi_1+c^\alpha_{00}\partial_\alpha\phi_2 \nonumber \\
&&\;\;\;=-(2\Gamma_{(01)11}-4\Gamma_{1110})\phi_0+(2\Gamma_{0011}-4\Gamma_{(01)01}-4\Gamma_{1100})\phi_1
+6\Gamma_{(01)00}\phi_2-2\Gamma_{0000}\phi_3, \label{c1} \\
&&c^0_{11}\partial_\tau\phi_1-2c^0_{01}\partial_\tau\phi_2+c^0_{00}\partial_\tau \phi_3 + c^\alpha_{11}\partial_\alpha\phi_1-2c^\alpha_{01}\partial_\alpha\phi_2+c^\alpha_{00}\partial_\alpha\phi_3 \nonumber \\
&&\;\;\;=\Gamma_{1111}\phi_0-(4\Gamma_{(01)11}-2\Gamma_{1101})\phi_1+3(\Gamma_{0011}-\Gamma_{1100})\phi_2-(2\Gamma_{0001}-4\Gamma_{(01)00})\phi_3\nonumber\\
&&\;\;\;\;\;\;-\Gamma_{0000}\phi_4, \label{c2}\\
&&c^0_{11}\partial_\tau\phi_2-2c^0_{01}\partial_\tau\phi_3+c^0_{00}\partial_\tau \phi_4 + c^\alpha_{11}\partial_\alpha\phi_2-2c^\alpha_{01}\partial_\alpha\phi_3+c^\alpha_{00}\partial_\alpha\phi_4 \nonumber \\
&&\;\;\;=2\Gamma_{1111}\phi_1-6\Gamma_{(01)11}\phi_2+(4\Gamma_{0011}+4\Gamma_{(01)01}-2\Gamma_{1100})\phi_3-(4\Gamma_{0001}-2\Gamma_{(01)00})\phi_4.
\label{c3}
\end{eqnarray}
\end{subequations}

\subsection{The initial data}
As pointed out in the introduction, only asymptotically Euclidean,
time symmetric, analytically conformally flat initial data 
will be considered in our discussion. A number of simplification arise
under these assumptions. In particular, around $i$, the conformal
factor $\Omega$ of the initial hypersurface can be written as,
\begin{equation}
\label{Omega_conformal_factor}
\Omega=\frac{\rho^2}{(1+\rho W)^2},
\end{equation}
where $W(i)=m/2$, $m$ the ADM mass of the initial hypersurface. The
function $W$ satisfies the Yamabe equation, which under our
assumptions reduces to the Laplace equation. Therefore $W$ is
harmonic, and thus can be written as,
\begin{equation}
\label{expansion_for_W}
W=\frac{m}{2}+ \sum_{p=1}^\infty \sum_{k=0}^{2p}\frac{1}{p!} w_{p,2p,k}T_{2p\phantom{k}p}^{\phantom{2p}k}\rho^p
\end{equation} 
where the coefficients $w_{p,2p,k}$, $p=1,2,\ldots$, $k=0,\ldots,2p$ complex
numbers satisfying the regularity condition
$\overline{w}_{p,2p,k}=(-1)^{i+k}w_{p,2p,2p-k}$ so that $W$ is a real valued
function. 

As mentioned in section 2, a crucial property of our set up based on the
properties of conformal geodesics is that it renders a conformal factor
$\Theta$ ---see formula (\ref{conformal_factor}) for the region of spacetime
under discussion. Furthermore, solving the conformal geodesic equations also
yields a 1-form $d_{ab}$, which appears in the propagation equations
(\ref{p6}) and (\ref{p7}). Under our assumptions of time symmetry and
conformal flatness it is given by,
\begin{equation}
\label{one_form}
d_{ab}=2\rho\left(\frac{x_{ab}-\rho^2D_{ab}}{(1+\rho W)^3}\right).
\end{equation}    

Once the function $\kappa$ of section 2 has been chosen, the initial data for
the conformal propagation equations (\ref{p1})-(\ref{p7}) is given by,
\begin{subequations}
\begin{eqnarray}
&& \Theta_{abcd}=-\frac{\kappa^2}{\Omega} D_{(ab}D_{cd)}\Omega, \label{initial1}\\
&& \phi_{abcd}=\frac{\kappa^3}{\Omega^2}D_{(ab}D_{cd)}\Omega, \label{initial2}\\
&& c^0_{ab}=0, \quad c^1_{ab}=\kappa x_{ab},  \label{initial3}\\
&& c^+=\frac{\kappa}{\rho}z_{ab}, \quad c^-=\frac{\kappa}{\rho}y_{ab},
 \label{initial4}\\
&& \xi_{abcd}=\sqrt{2}\left\{ \frac{\kappa}{2\rho}(\epsilon_{ac}
  x_{bd} + \epsilon_{bd} x_{ac})
  -\frac{1}{2\kappa}(\epsilon_{ac}D_{bd}\kappa+\epsilon_{bd}D_{ac}\kappa)\right\}, \label{initial5}\\
&& \chi_{(ab)cd}=0, \quad f_{ab}=D_{ab}\kappa. \label{initial6}
\end{eqnarray} 
\end{subequations}
where $D_{ab}$, the spinorial covariant derivative of the initial hypersurface
$S$ is given by,
\begin{equation}
D_{ab}\mu_{cd}=x_{ab}\partial_\rho\mu_{cd} +\frac{1}{\rho}z_{ab}X_+\mu_{cd} +\frac{1}{\rho}y_{ab}X_-\mu_{cd}   -\gamma_{ab\phantom{e}c}^{\phantom{ab}e}\mu_{ed} -\gamma_{ab\phantom{e}d}^{\phantom{ab}e}\mu_{ec},
\end{equation} 
where the flat connection coefficients $\gamma_{abcd}$ are given by,
\begin{equation}
\gamma_{abcd}=\frac{1}{2\rho}(\epsilon_{ac}x_{bd}+\epsilon_{bd}x_{ac}),
\end{equation}
for a given differentiable spinorial function $\mu_{ab}$. 

\section{The transport equations}
The equations
(\ref{p1})-(\ref{p7}) can be concisely written in the form,
\begin{equation}
\label{v_equations}
\partial_\tau v = Kv +Q(v,v)+ L\phi,
\end{equation}
where $K$ and $Q$ are respectively a linear and a quadratic function with
constant coefficients, whereas $L$ is a linear function depending
on the coordinates via $\Theta$, $\partial_\tau\Theta$ and
$d_{ab}$. For the Bianchi propagation equations (\ref{b0})-(\ref{b4}) one can
write,
\begin{equation}
\label{b_equations}
\sqrt{2}E\partial_\tau\phi + A^{ab}c^\mu_{ab}\partial_\mu\phi=B(\Gamma_{abcd})\phi, 
\end{equation}
where now, $E$ denotes the $(5\times 5)$ unit matrix, $A^{ab}c^\mu_{ab}$
are $(5\times 5)$ matrices, and $B(\Gamma_{abcd})$ is a linear
$(5\times 5)$-matrix valued function of the connection coefficients
$\Gamma_{abcd}$. Similarly, the Bianchi constraint equations
(\ref{c1})-(\ref{c3}) can be written as,
\begin{equation}
F^{ab}c^\mu_{ab} \partial_\mu\phi = H(\Gamma_{abcd}),
\end{equation}
where now $F^{ab}c^\mu_{ab}$ denotes a $(3\times 5)$ matrix, and
$H(\Gamma_{abcd})$ denotes a $(3\times 5)$ matrix valued function  of the
connection. 

In the sequel, given an unknown $u$ we will write, $u^{(0)}=u|_I$. The objects
$\Theta$, $\partial_\tau\Theta$ and $d_{ab}$ from which $L$ is constructed,
vanish on $I$. Thus, $L^{(0)}=L|_I$, and consequently the equations
(\ref{v_equations}) and (\ref{b_equations}) decouple from each other.  The
system of equations for the $v$ unknowns, equation (\ref{v_equations}), turns
out to be an interior system upon evaluation on the cylinder at spatial
infinity. Its initial data can be read from the restriction of the initial
data (\ref{initial1})-(\ref{initial6}) to $I$. It can be seen that this
restriction, irrespectively of the choice of the function $\kappa$ coincides
with the initial data of Minkowski spacetime. With this information in hand,
the system for $v^{(0)}$ can be readily solved yielding,
\begin{subequations}
\begin{eqnarray}
&& \Theta^{(0)}_{abcd}=0, \quad \chi^{(0)}_{(ab)cd}=0, \quad f^{(0)}_{ab}=0, \quad
\xi^{(0)}_{abcd}=0, \label{solution_order_0a}\\
&& (c^0_{ab})^{(0)}=-\tau x_{ab}, \quad (c^1_{ab})^{(0)}=0, \quad
(c^-_{ab})^{(0)}=y_{ab}, \quad (c^+_{ab})^{(0)}=z_{ab}. \label{solution_order_0b}
\end{eqnarray}
\end{subequations}
From this solution it follows that the matrix  $A^{ab}c^1_{ab}$ in
the system (\ref{b_equations}) satisfies,
\begin{equation}
A^{ab}c^1_{ab}|_I=0.
\end{equation}
This particular result will be crucial in our later discussion. As a
consequence of it, the system (\ref{b_equations}) implies another interior
system on I, as no $\rho$-derivatives will arise upon evaluation on $I$. It
can be solved giving,
\begin{equation}
\phi^{(0)}_{abcd}=-6m\epsilon^2_{abcd}.
\end{equation}

From the latter discussion that the cylinder at spatial infinity $I$
is a characteristic of the system
(\ref{v_equations})-(\ref{b_equations}). Furthermore, because of the
fact that the whole system of conformal field equations reduces to an
interior system on $I$ ---something that does not happen with normal
characteristics--- we call it a \emph{total characteristic}. 

The idea of interior systems previously discussed can be generalised by
applying $p$ times $\partial_\rho$ to the equations (\ref{v_equations}) and
(\ref{b_equations}) and then evaluating on $I$. In this way one obtains a
hierarchy of interior systems for the unknowns $u^{(p)}=\partial_\rho u|_I$.
These quantities can be used to construct formal expansions of the form,
\begin{equation}
u=\sum_{p=0} \frac{1}{p!}u^{(p)}\rho^p
\end{equation}
for the field quantities. The resulting equations, which will be
referred generically to as \emph{transport equations}, are of the
form,
\begin{eqnarray}
&&\partial_\tau v^{(p)} = Kv^{(p)}
+Q(v^{(0)},v^{(p)})+Q(v^{(p)},v^{(0)}) \nonumber \\
&&\phantom{XXXXXXXX}+\sum_{j=1}^{p-1}\left(
  Q(v^{(j)},v^{(p-j)})+ L^{(j)}\phi^{(p-j)}\right) +
L^{(p)}\phi^{(0)}, \label{v_transport}
\end{eqnarray} 
which correspond to the propagation equations of the $v$ unknowns,
equations (\ref{p1})-(\ref{p7}). From the Bianchi propagation
equations (\ref{b0})-(\ref{b4}) one gets,
\begin{eqnarray}
&&\left( \sqrt{2}E +A^{ab}(c^0_{ab})^{(0)}\right)\partial_\tau^{(p)} +
A^{ab}(c_{ab}^C)^{(0)}\partial_C\phi^{(p)}= B(\Gamma^{(0)}_{abcd})\phi^{(p)}
\nonumber \\
&&\phantom{XXXXXXXX}+\sum_{j=1}^p
\left(\begin{array}{c} p \\ j
  \end{array}\right)\left(B(\Gamma_{abcd}^{(j)})\phi^{(p-j)}-A^{ab}(c^\mu_{ab})^{(j)}\partial_\mu
 \phi^{(p-j)}\right), \label{b_transport}
\end{eqnarray}
where $C=\pm$. Similarly, from the Bianchi constraint equations
(\ref{c1})-(\ref{c3}) one obtains,
\begin{eqnarray}
&&F^{ab}(c^0_{ab})^{(0)}\partial_\tau\phi^{(p)}+F^{ab}(c^C_{ab})^{(0)}\partial_C\phi^{(p)}=H(\Gamma^{(0)}_{abcd})
  \nonumber \\
&&\phantom{XXXXXXXX}+\sum_{j=1}^p
\left(\begin{array}{c} p \\ j
  \end{array}\right)
 \left(
   H(\Gamma_{abcd}^{(j)})\phi^{(p-j)}-F^{ab}(c^\mu_{ab})^{(j)}\partial_\mu\phi^{(p-j)}\right). \label{c_transport}
\end{eqnarray}
The systems (\ref{v_transport}) and (\ref{b_transport}) can be
regarded as systems for the unknowns $v^{(p)}$ and $\phi^{(p)}$ if the
lower order quantities $v^{(j)}$ and $\phi^{(j)}$, $0\leq k \leq p-1$
are known. The two systems are decoupled from each other, and
accordingly one would firstly solve the system (\ref{v_transport}) and
then feed its solution into the system (\ref{b_transport}) which now
could in turn be solved.
 
\begin{figure}[t]
\centering
\begin{tabular}{cc}
\includegraphics[width=.4\textwidth]{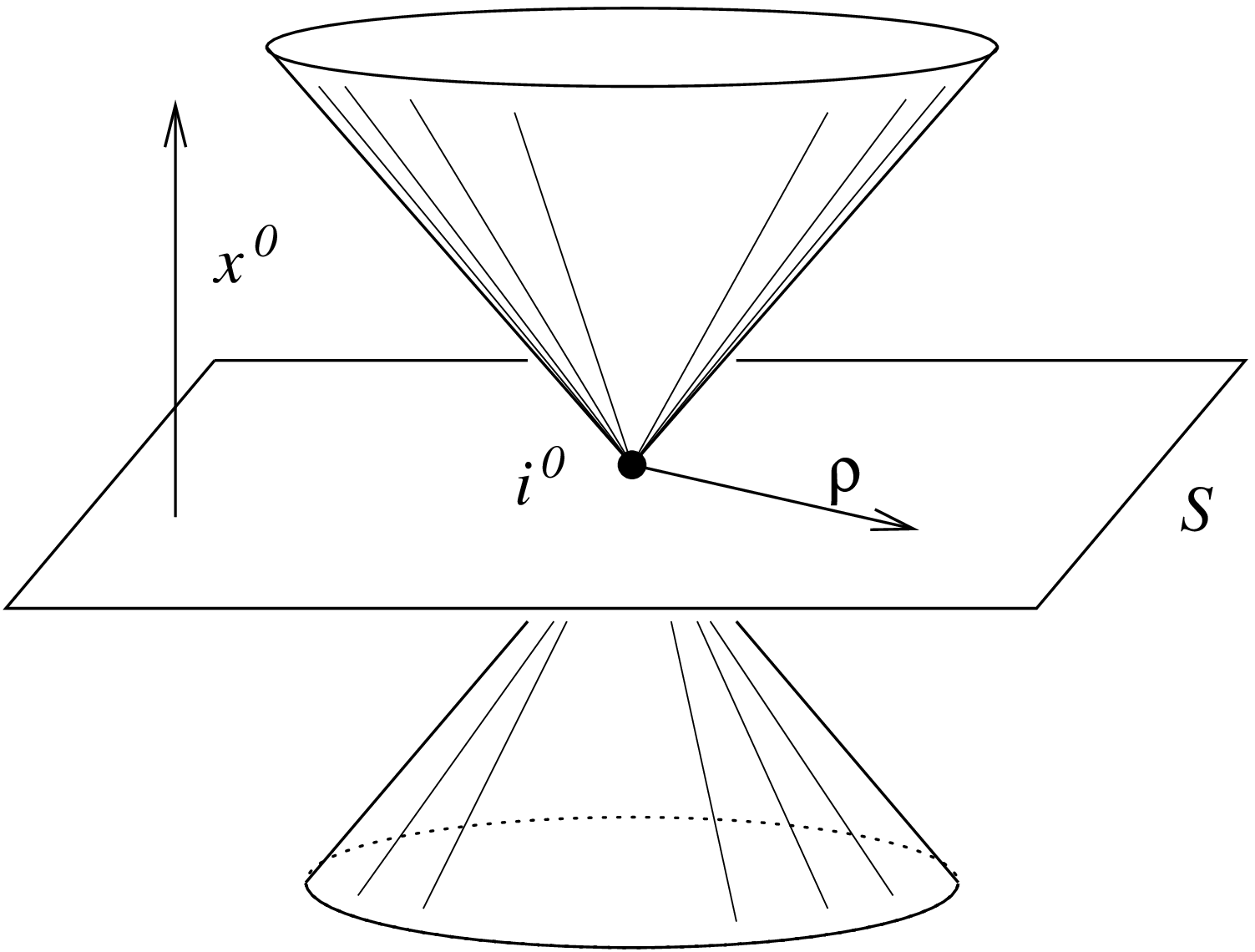} &
\includegraphics[width=.4\textwidth]{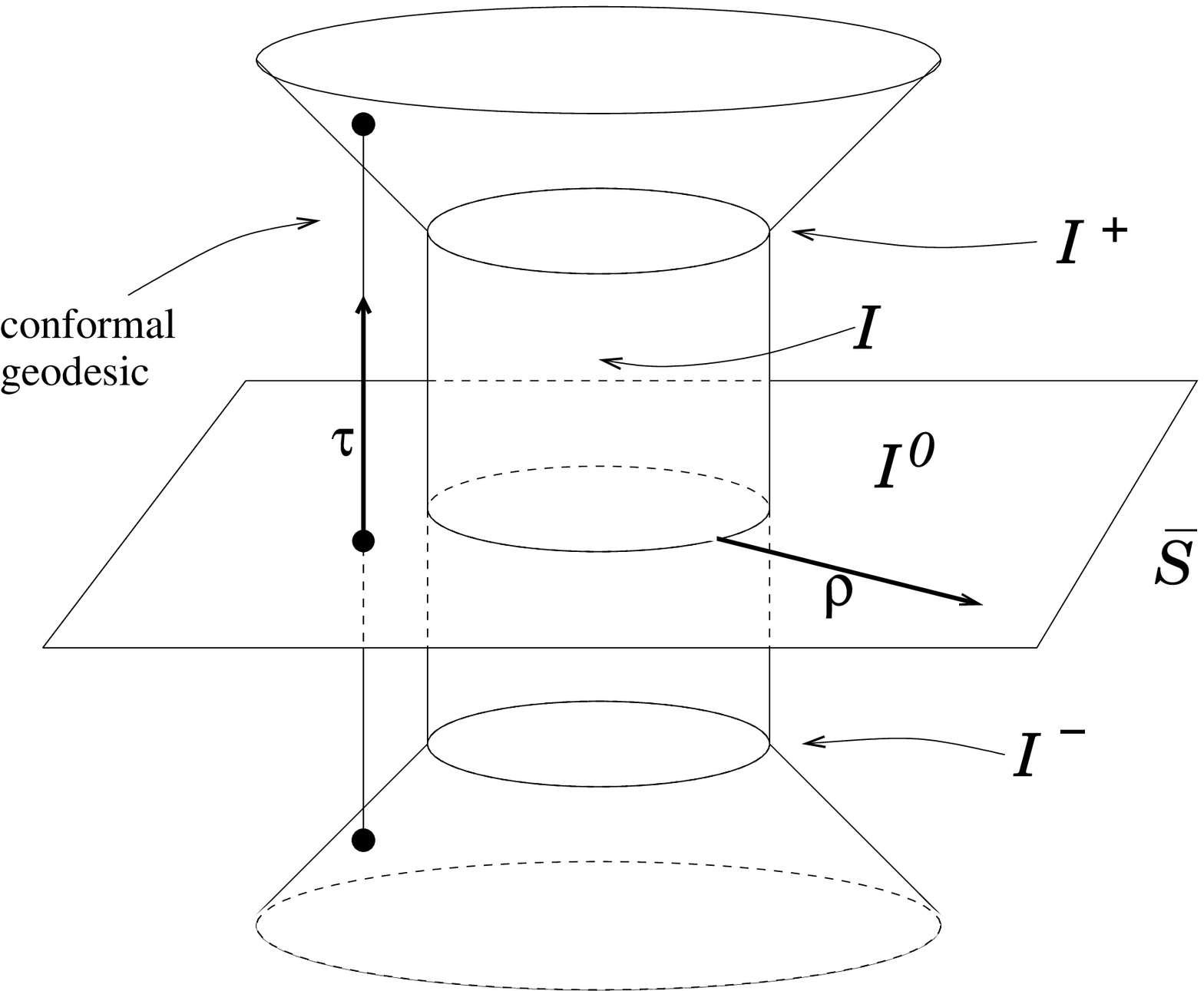}
\end{tabular}
\put(-25,60){$\scri^+$}
\put(-25,-60){$\scri^-$}
\put(-215,50){$\scri^+$}
\put(-215,-50){$\scri^-$}
\caption[]{Spacetime close to spatial and null infinities: to the left
  the standard representation of spatial infinity as a point $i^0$; to
  the right the representation where spatial infinity is envisaged as a
  cylinder.}
\label{old_vs_new}
\end{figure}

Because of (\ref{solution_order_0a}) and
(\ref{solution_order_0b}), the matrix accompaigning the $\partial_\tau$
derivative in the system (\ref{b_transport}) is given by,
\begin{equation}
\left( \sqrt{2}E
  +A^{ab}(c^0_{ab})^{(0)}\right)=\sqrt{2}\,\mbox{diag}(1+\tau,1,1,1,1-\tau) \label{principal_part}
\end{equation}
As a consequence of this, the symbol of the system looses rank at $\tau=\pm 1$
---the system degenerates there. The points $\tau=\pm 1$ corresponds precisely
to the sets $I^\pm$ ---cfr. (\ref{degenerate_sets})--- the sets where ``null
infinity touches spatial infinity''. It is exactly this particular feature of
the field equations that forces us to undertake a complicate and detailed
analysis of the system (\ref{b_equations}). From an heuristic point of view
the degeneracy at the sets $I^\pm$ can be understood as a consequence of the
change of behaviour of the conformal boundary of the spacetime with regard to
the conformal field equations: the cylinder at spatial infinity $I$ is a total
characteristic, while the $\scri^\pm$ are ``only'' normal characteristics
---i.e. only subsets of the field equations reduce to interior systems on
either $\scri^+$ or $\scri^-$. Now, standard theory of symmetric
hyperbolic systems guarantees, for a given order $p$, the existence of
solutions to the joint system (\ref{v_transport})-(\ref{b_transport})
for any subset of $I$. However, for the sets $I^\pm$ the degeneracy
implied by the equation (\ref{principal_part}) the usual energy
estimates provide no information precisely at the point one is
interested most---a similar phenomenon occurs with the
original system (\ref{v_equations})-(\ref{b_equations}). Thus, one
needs to devise non-standard methods in order to address existence
issues ---see for example the discussions in \cite{Fri02b,Val02a}.

So far, the function $\kappa$ introduced in equation
(\ref{conformal_factor}) has been required to be of the form
$\kappa=\rho\kappa'$, with $\kappa'$ analytic an such that
$\kappa'(i)=1$, but otherwise it has remained unspecified \footnote{It
  is noted that the simple choice $\kappa=1$ would lead to the
  standard representation of spatial infinity as a point ---see
  figure. The requirement of $\kappa$ being of the form
  $\kappa=\rho\kappa'$ ensures that spatial infinity is blown up to a
  cylinder. See figure \ref{old_vs_new}}. Two choices consistent with
the requirement $\kappa=\rho \kappa'$ will be considered here. The
first, $\kappa=\rho$ is the simplest non-trivial one. For this choice
$\scri^+$ in a neighbourhood of $I^+$ is concave, while $\scri^-$ in a
neighbourhood of $I^-$ would be convex. The choice $\kappa=\rho$ has
the virtue of rendering the simplest possible analytic expressions,
both for the initial data (\ref{initial1})-(\ref{initial6}) and the
solution of the transport equations
(\ref{v_transport})-(\ref{b_transport}). Unfortunately, it is hard to
attach to it some geometrical significance other that its simplicity.
The other choice to be considered is $\kappa=\omega$. This choice is
fine as under our assumptions $\omega=\rho+\O(\rho)$. With this
choice,
\begin{equation}
\Theta=\omega^{-1} \Omega\left(1-\tau^2\right),
\end{equation}
so that $\scri^\pm$ near $I^\pm$ are described by the hypersurfaces
$\tau=\pm 1$, $\rho>0$ respectively: null infinity will be composed of
two parallel planes, formally similar to the case of Minkowski ---see
\cite{Val02a} \footnote{This similarity is in some aspects deceiving,
  as generically when $m\neq 0$ the generators of null infinity,
  although confined to the planes $\tau=\pm$, are bent and may rotate
  a spin frame that is parallelly transported along it}. Consequently,
the system of conformal field equations
(\ref{v_equations})-(\ref{b_equations}) degenerate not only on $I^\pm$
but also on $\scri^\pm$. Thus, the choice $\kappa=\omega$ has more
geometrical and analytic relevance. As a drawback it renders more
complicated analytic expressions.

\begin{figure}[t]
\centering
\begin{tabular}{cc}
\includegraphics[width=.4\textwidth]{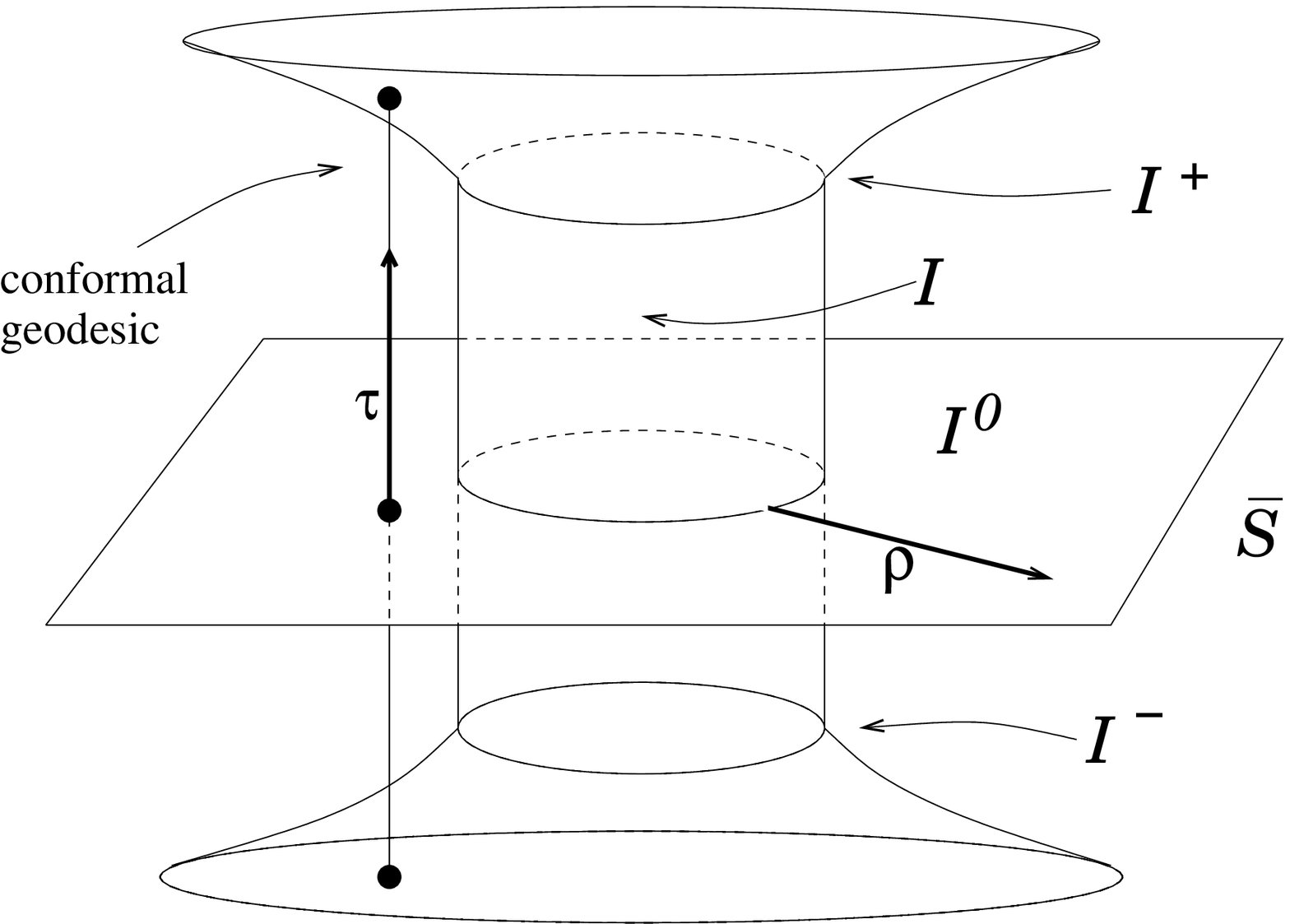} &
\includegraphics[width=.4\textwidth]{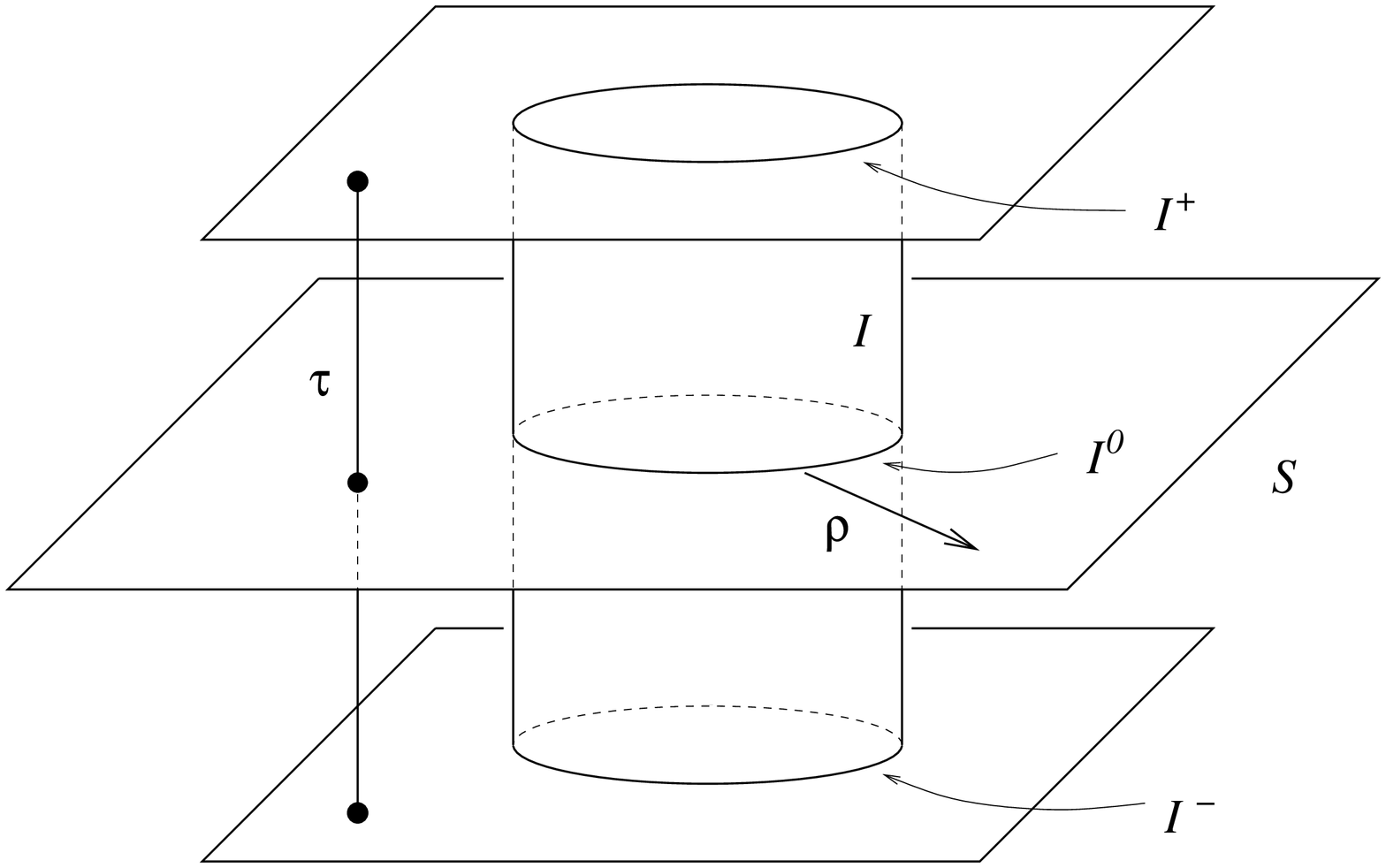}
\end{tabular}
\put(-25,50){$\scri^+$}
\put(-25,-30){$\scri^-$}
\put(-200,55){$\scri^+$}
\put(-200,-50){$\scri^-$}
\caption[]{Effect of the choice of the function $\kappa$ on the
  representation of null infinity near spatial infinity. To the left
  with the choice $\kappa=\rho$; to the right the choice
  $\kappa=\omega$ so that null infinity corresponds to the
  hypersurfaces $\tau=\pm 1$.}
\label{characteristic}
\end{figure}

For future use, we note the following result on the expansion types of
the diverse unknowns appearing in the transport equations
(\ref{v_transport}) and (\ref{b_transport}). Its proof comes from
inspection \cite{Fri98a}.
\begin{lemma}
\label{lemma:expansion_types}
The functions $\left( c^1_{ab}-\rho x_{ab}\right)^{(p)}$, $v^{(p)}$,
$\phi^{(p)}$ $p=1,2,\dots$, are of expansion type $p-2$, $p-1$, and
$p$ respectively.
\end{lemma}

\subsection{A first analysis of the transport equations}

A detailed first analysis of the structure of the
solutions of the transport equations have been given in
\cite{Fri98a}. Because of its relevance for the purposes of the
present article, and in order to fully motivate our later discussion,
we briefly proceed to review it. 

As seen in the previous section, the transport equations corresponding
to the $v^{(p)}$ quantities, equation (\ref{v_transport}) are fully
regular, thus in a first analysis one could just focus on the
transport equations arising from the Bianchi identities, equation
(\ref{b_transport}). The system can be written as,
\begin{subequations}
\begin{eqnarray}
&&(1+\tau)\partial_\tau\phi_0^{(p)} +
X_+\phi_1^{(p)}-(p-2)\phi_0^{(p)}=R_0^{(p)}, \label{T0} \\
&&\partial_\tau\phi_1^{(p)}
+\frac{1}{2}(X_+\phi_2^{(p)}+X_-\phi_0^{(p)})+\phi_1^{(p)}=R_1^{(p)},
\label{T1}\\
&&\partial_\tau\phi_2^{(p)}
+\frac{1}{2}(X_+\phi_3^{(p)}+X_-\phi_1^{(p)})=R_2^{(p)},
\label{T2}\\
&&\partial_\tau\phi_3^{(p)}
+\frac{1}{2}(X_+\phi_4^{(p)}+X_-\phi_2^{(p)})-\phi_3^{(p)}=R_3^{(p)},
\label{T3}\\
&&(1-\tau)\partial_\tau\phi_4^{(p)} +
X_-\phi_3^{(p)}+(p-2)\phi_4^{(p)}=R_4^{(p)}\label{T4}.
\end{eqnarray}
\end{subequations}
We will also make use of the transport equation arising from the
Bianchi transport equations, equation (\ref{c_transport}). These we
write as,
\begin{subequations}
\begin{eqnarray}
&& \tau\partial_\tau \phi_1^{(p)}
+\frac{1}{2}(X_+\phi_2^{(p)}-X_-\phi_0^{(p)})-p\phi_1^{(p)}=S_1^{(p)},
\label{T5}\\
&&\tau\partial_\tau \phi_2^{(p)}
+\frac{1}{2}(X_+\phi_3^{(p)}-X_-\phi_1^{(p)})-p\phi_2^{(p)}=S_2^{(p)},
\label{T6}\\
&&\tau\partial_\tau \phi_3^{(p)}
+\frac{1}{2}(X_+\phi_4^{(p)}-X_-\phi_2^{(p)})-p\phi_3^{(p)}=S_3^{(p)} \label{T7}.
\end{eqnarray}
\end{subequations}
The quantities $R^{(p)}_j$, $j=0,\ldots,4$ and $S^{(p)}_i$, $i=1,2,3$,
are constructed from $v^{(k)}$ unknowns with $0\leq k \leq p$, and
from $\phi^{(k)}$ quantities with $0\leq k \leq p-1$. All these
quantities are assumed to be known. In this first
analysis their detailed structure will no be crucial. It will suffice
to know that they are of expansion type $p-1$. On the other hand,
consequently with lemma \ref{lemma:expansion_types} we write, 
\begin{equation}
\label{phi_decomposed}
\phi^{(p)}_j=\sum_{q=|2-j|}^p \sum_{k=0}^{2q} a_{j,p;2q,k}T_{2q\phantom{k}q-2+j}^{\phantom{2q}k},
\end{equation}
where the coefficients $a_{j,p;2q,k}=a_{j,p;2q,k}(\tau)$ are complex
valued functions. The expression (\ref{phi_decomposed}) effectively
introduces a decomposition into spherical harmonics for the equations
(\ref{T0})-(\ref{T7}). In this way one has only to deal with ordinary
differential equations. By taking linear combinations of equations
(\ref{T1}) and (\ref{T5}), (\ref{T2}) and (\ref{T6}), (\ref{T3}) and
(\ref{T7}) it can be readily seen
that the coefficients $a_{1,p;2q,k}$, $a_{2,p;2q,k}$, $a_{3,p;2q,k}$
can be determined once the coefficients  $a_{0,p;2q,k}$ and
$a_{4,p;2q,k}$ are known. These last two coefficients can be seen to
satisfy the following equations,
\begin{subequations}
\begin{eqnarray}
&&(1+\tau)f(\tau) a'_{0,p;2q,k} +g(\tau)a_{0,p;2q,k} +h(\tau)a_{4,p;2q,k}= T_0, \label{mess_1}\\
&&-(1-\tau)f(-\tau) a'_{4,p;2q,k} +g(-\tau) a_{4,p;2q,k} +h(\tau) a_{0,p;2q,k}=T_4 \label{mess_2},
\end{eqnarray}
\end{subequations}
where
\begin{eqnarray*}
&&f(\tau)=2(p+1)(p-1)-(q-1)(q+2)(1-\tau^2),\\
&&g(\tau)=-(p-2)f(\tau)+(q-1)(q+2)\left((p+1)(1+\tau)-(1+\tau)^2-\frac{1}{4p}q(q+1)(1-\tau)(1+\tau)^2\right),
\\
&&h(\tau)=\frac{1}{4p}(q-1)q(q+1)(q+2)(1-\tau)^3,
\end{eqnarray*} 
and $T_0$, $T_4$, depend on the coefficients $a_{j',p';q',k'}$ with
$p<p'$. They vanish, in particular, for $q=p$ as a consequence of
$R^{(p)}_j$ and $S^{(p)}_i$ being of expansion type $p-1$. In order to
solve the system (\ref{mess_1})-(\ref{mess_2}), one can try to
construct a fundamental matrix of the homogeneous system. Remarkably,
the homogeneous versions of equations (\ref{mess_1}) and
(\ref{mess_2}) can be decoupled to obtain two Jacobi equations (second
order). This class of equations have been studied extensively in the
literature.

For $0\leq q \leq p-1$ one gets for the system
(\ref{mess_1})-(\ref{mess_2}) a fundamental matrix,
\begin{equation}
X_{p;2q,k}=\left(
\begin{array}{cc}
\check{Q}_{p;2q,k}(\tau) & (-1)^q\hat{Q}_{p;2q,k}(\tau) \\
(-1)^q\hat{Q}_{p;2q,k}(-\tau) & \check{Q}_{p;2q,k}(-\tau) \\
\end{array}
\right),
\end{equation}
where 
\begin{equation}
\check{Q}_{p;2q,k}(\tau)=\left(\frac{1-\tau}{2}\right)^{p+2}P_{q-2}^{(p+2,2-p)}(\tau),
\end{equation}
and 
\begin{equation}
\hat{Q}_{p;2q,k}(\tau)=\left(\frac{1+\tau}{2}\right)^{p-2}P_{q+2}^{(-p-2,p-2)}(\tau),
\end{equation}
with $P_n^{(\alpha,\beta)}(\tau)$ the generalised Jacobi
polynomials,
\begin{equation}
P_n^{(\alpha,\beta)}(\tau)=\sum_{\nu=0}^{n} \binom{n+\alpha}{n-\nu}
\binom{n+\beta}{\nu}\left(\frac{\tau-1}{2}\right)^\nu \left(\frac{\tau+1}{2}\right)^{n-\nu}.
\end{equation}
It is noted by passing that the fundamental matrix $X_{p;2q,k}$ for
given $p$ and $q$ does not depend on $k$. If $p=q$ the story is other.
In this case the fundamental matrix is of the form,
\begin{equation}
X_{p;2p,k}=\left( \begin{array}{cc}
c_{11}\check{Q}_{p;2p,k}(\tau) & c_{12}\hat{Q}_{p;2p,k}(\tau) \\
c_{21}\hat{Q}_{p;2p,k}(-\tau) & c_{22}\check{Q}_{p;2p,k}(-\tau) \\
\end{array}
\right),
\end{equation}
where,
\begin{eqnarray}
&&\check{Q}_{p;2p,k}(\tau)=\left(\frac{1-\tau}{2}\right)^{p+2}\left(\frac{1+\tau}{2}\right)^{p-2}
\\
&&\hat{Q}_{p;2p,k}(\tau)=\int_0^\tau\frac{ds}{(1+s)^{p-1}(1-s)^{p+3}},
\end{eqnarray}
and $c_{11}$, $c_{12}$, $c_{21}$ and $c_{22}$ are numerical
constants. A simple induction argument shows that,
\begin{eqnarray*}
&&\int_0^\tau\frac{ds}{(1+ s)^{p-1}(1-
  s)^{p+3}}=A_{*}\ln(1-\tau)+\frac{A_{p+ 2}}{(1-\tau)^{p+
    2}}+\cdots+\frac{A_{1}}{(1-\tau)}+A_0
\nonumber \\
&&\phantom{\int_0^\tau\frac{ds}{(1+s)^{p-1}(1-s)^{p+3}}=}
+B_{*}\ln(1+\tau)+\frac{B_{p- 2}}{(1+\tau)^{p- 2}}+\cdots+\frac{B_{1}}{(1+\tau)},
\end{eqnarray*}
where the $A$'s and the $B$'s are some numerical coefficients. They
precise values are not relevant for our purposes. It follows from the
latter discussion that any solution for the sectors with $p=q$ will
contain logarithmic divergences at $\tau=\pm 1$ unless
$a_{0,p;2p,k}(0)=a_{4,p;2p,k}(0)$. 
Quite remarkably, for time symmetric initial data this quite involved
and technical-looking condition can be reexpressed in terms of a very
geometric and appealing condition on the Cotton-Bach tensor of the
initial data at spatial infinity \cite{Fri98a}. One has the following lemma: 
\begin{lemma}
For time symmetric initial data the following two conditions are equivalent,
\begin{itemize}
\item[(i)] 
\begin{equation}
a_{0,p;2p,k}(0)=a_{4,p;2p,k}(0)
\end{equation} 
with $p\geq 2$ and $k=0,\ldots,2p$,
\item[(ii)]
\begin{equation}
\label{regcond} 
D_{(a_sb_s}\cdots D_{a_1b_1} b_{abcd)}(i)=0,
\end{equation}
with $s=0,1,\ldots$\;\;.
\end{itemize} 
\end{lemma}
From the above discussion it should be clear that the condition
(\ref{regcond}) is a necessary requirement in order to for the
solutions $\phi_j^{(p)}$ of the transport equations
(\ref{b_transport}) are smooth. Therefore we refer to condition
(\ref{regcond}) as to a \emph{regularity condition}, and we have the
following very suggestive result, 
\begin{theorem}[Friedrich, 1997]
The solutions $u^{(p)}$ of the transport equations extend smoothly to
the sets $I^\pm$ only if the condition
\[
D_{(a_sb_s}\cdots D_{a_1b_1} b_{abcd)}(i)=0, \quad s=0,1,\ldots
\]
is satisfied at all orders $s$. It is is not satisfied for some $s$,
the solution develops logarithmic singularities at $I^\pm$. 
\end{theorem}

It should be emphasized that the consideration leading to this last
theorem have hardly make use of the inhomogeneous terms of equations
(\ref{mess_1}) and (\ref{mess_2}). The system
(\ref{mess_1})-(\ref{mess_2}) can be in matricial form as,
\begin{equation}
\label{reduced_system}
A_{p;2q}(\tau) y'_{p;2q,k}(\tau)+ B(\tau)y_{p;2q,k}(\tau)= t_{p;2q,k},
\end{equation} 
where $A_{p;2q}$ and $B_{p;2q}$ are $2\times 2$, and $t_{p;2q,k}$ is a
column vector such that $t_{p;2p,k}=0$. The the solution of system
given in terms of the fundamental matrix is then,
\begin{equation}
\label{reduced_solution}
y_{p;2q,k}(\tau)=X_{p;2q,k}(\tau)X^{-1}_{p;2q,k}(0)y_{p;2q,k}(0)+\int_0^\tau
X_{p;2q,k}^{-1}(s)\;A_{p;2q}^{-1}(s)\;t_{p;2q,k}(s) ds.
\end{equation}
Assume that the regularity condition (\ref{regcond}) holds. Even in
this case the integrand 
$X^{-1}_{p;2q}(\tau)A^{-1}_{p;2q}(\tau)\;t_{p;2q,k}(\tau) $ in
formula (\ref{reduced_solution}) contains poles at $\tau=\pm 1$ and
possibly also outside the interval $[-1,1]$. Therefore, unless some
remarkable property of the conformal field equations comes into stage,
the solution vector $y_{p;2q,k}$ will contain some logarithmic terms
and consequently also the coefficients $\phi_j^{(p)}$.

The algebraic structure of the integral in (\ref{reduced_solution}) is
too complicated to analyse directly without resorting to some further
(still unknown) structure of the field equations. In order to shed
some light into this direction, Friedrich \& K\'{a}nn\'{a}r have
performed an analysis of the solutions of the transport equations up
to order $p=3$ under the assumption that the initial data satisfies
the condition (\ref{regcond}). Quite remarkably they found that no
logarithms appeared in the solutions. This meant that somehow, the
simple poles in (\ref{reduced_solution}) cancel out up to the order
under consideration.

At this point it is important to mention that due to the gauge
conditions used in the whole construction, we know that the
logarithmic singularities occurring in the solutions to the transport
equations are associated with the conformal structure, and are not
artifacts of a choice of gauge. Note that the condition equation
(\ref{regcond}) is conformally invariant. It allows us to identify
obstructions to the smoothness of null infinity explicitly in terms of
the initial data. Also, because of the hyperbolic nature of the
equations suggest that if logarithmic terms are present in the
solutions of the transport equations, then it is very likely these
will propagate along $\scri$.

\section{Solving the transport equations given time symmetric,
  conformally flat initial data}

On simplicity and aesthetical grounds it is natural to wonder whether
the regularity condition, equation (\ref{regcond}), is the only
requirement one has to impose on the initial data in order to obtain
solutions to the transport equations which are smooth ---see for
example the discussion in \cite{Fri02a}. Before trying to prove some
statement along this lines, it is of clear interest to calculate the
some further order in the expansions. The rationale behind it being
firstly to verify whether the conjecture still holds, and secondly
to try to find some patterns in the solutions that one could exploit
in an eventual abstract proof. In order to simplify the calculations,
we have chosen to restrict our attention to those time symmetric
initial data sets which are conformally flat. These satisfy the
regularity condition (\ref{regcond}) in a trivial way. Thus, they
represent the simplest (non-trivial) class of initial data sets one
can look at. Their simplicity is somehow deceiving, and should not be
regarded as a drawback on the kind of insight that can be gained
through them. A great deal about the solutions of the Einstein field
equations has been learned from the analysis of this class of initial
data ---see e.g. \cite{BriLin63,Mis63,Val02b}.

The already ``big'' transport equation systems
(\ref{v_transport})-(\ref{b_transport}) do not give an appropriate
dimension of the computational difficulties one has to face if one is
to take the expansions carried in \cite{FriKan00} to even higher
orders. However, the calculations one has to carry are suitable to a
treatment using a computer algebra system. In order to analyse with
ease the solutions of the transport equations some scripts in the
computer algebra system {\tt Maple V} have been written. The reader is
remitted to the appendix for further details on the implementation of
the calculations on this system.

Because of the largeness of the expressions contained inthe solutions,
we have opted to provide a description of the qualitative features of
the expansions rather than a complete list of all the terms
calculated. In particular, attention will be focused on the solutions
to the transport equations arising from the Bianchi identities, the
functions $\phi_j^{(p)}$. It should be emphasized that this does not
mean that the solutions to the $v$ unknown transport equations are not
important. They are also crucial: a tiny mistake in the calculation of
their solutions would destroy the whole structure of the solutions.
However, as the discussion in the previous section has pointed out,
the logarithmic terms that destroy the smoothness of null infinity
appear firstly in the components of the Weyl spinor.

In order to perform our expansions a number of assumptions have been
made. We list them here for the purposes of a quick reference.  

\begin{assumptions}
It will be assumed that:
\begin{itemize}
\item[(i)] the initial data set is asymptotically Euclidean and  time symmetric.

\item[ii)] in a neighbourhood $B_a(i)$ of $i$ the initial
  data is assumed to be conformally flat. The function $W$ appearing
  in the conformal factor of the initial hypersurface ---see equation
  (\ref{Omega_conformal_factor})--- is a solution of the Laplace equation admitting in $B_a(i)$ a decomposition of the form,
\begin{equation}
W= \frac{1}{2}m + \sum_{i=1}^8 \frac{1}{i!}W_i \rho^i +\O(\rho^9),
\end{equation}
where 
\begin{equation}
W_i=\sum_{k=0}^{2i} w_{i,2i,k} T_{2i\phantom{k}i}^{\phantom{2i}k},
\end{equation}
with the coefficients $w_{i,2i,k}$, $i=1,\ldots, 7$, $k=0,\dots,2i$
complex numbers satisfying
$\overline{w}_{i,2i,k}=(-1)^{i+k}w_{i,2i,2i-k}$ so that $W$ is a real
valued function. This is in consistency with the decomposition given
in equation (\ref{expansion_for_W}).

\item[(iii)] Likewise, the components of the vector unknowns $v^{(p)}$
  and $\phi^{(p)}$ admit on $I$ expansions in terms of
  $T_{j\phantom{k}l}^{\phantom{j}k}$ functions consistent with lemma
  \ref{lemma:expansion_types}.

\item[(iv)] The two following choices of the function $\kappa$ ---see
  equation (\ref{conformal_factor})---  will be considered:
\begin{eqnarray*}
&& \kappa_1=\rho, \\
&& \kappa_2=\omega.
\end{eqnarray*}
\end{itemize}
\end{assumptions}

The result of the calculations under these assumptions are now described.

\subsection{The orders $p=0,\ldots,4$.}

Firstly, calculations for the orders $p=1$, $p=2$ and $p=3$ were
undertaken. The results are in complete agreement with those given by
Friedrich \& K\'{a}nn\'{a}r when reduced to the case of conformally
flat initial data.

The solutions at order $p=0$ can be schematically written as,
\begin{subequations}
\begin{eqnarray}
&& \phi_0^{(0)}=0,  \\
&& \phi_1^{(0)}=0,  \\
&& \phi_2^{(0)}=-m, \\
&& \phi_3^{(0)}=0,  \\
&& \phi_4^{(0)}=0. 
\end{eqnarray}
\end{subequations}
independently of the choice of $\kappa$. At order $p=1$ one has,
\begin{subequations}
\begin{eqnarray}
&& \phi_0^{(1)}=0,  \\
&& \phi_1^{(1)}=-3\,X_+W_1(1-\tau)^2, \\
&& \phi_2^{(1)}=m^2(\frac{1}{2}\tau^4-3\tau^2)+6\,W_1(\tau^2-1) \\
&& \phi_3^{(1)}=3\,X_-W_1(1+\tau)^2, \\
&& \phi_4^{(1)}=0. 
\end{eqnarray}
when $\kappa=\rho$. The expressions for $\phi_0^{(1)}$,
$\phi_1^{(1)}$, $\phi_3^{(1)}$ and $\phi_4^{(1)}$ for the choice
$\kappa=\omega$ are similar. That of $\phi_2^{(1)}$ is given by,
\begin{equation}
\phi_2^{(1)}=m^2(\frac{1}{2}\tau^4-3\tau^2-\frac{3}{2})+6\,W_1(\tau^2-1).
\end{equation}
\end{subequations}
At order $p=2$ one has,
\begin{subequations}
\begin{eqnarray}
&& \phi_0^{(2)}=f_1(\tau)X_+X_+W_2  , \\
&& \phi_1^{(2)}=f_2(\tau)mX_+W_1 + f_3(\tau)X_+W_2, \\
&& \phi_2^{(2)}=f_4(\tau)m^3+f_5(\tau)mW_1+f_6(\tau)W_2 ,\\
&& \phi_3^{(2)}=-f_2(-\tau)mX_-W_1 - f_3(-\tau)X_-W_2, \\
&& \phi_4^{(2)}=f_1(-\tau)X_-X_-W_2 , 
\end{eqnarray}
\end{subequations}
where $f_i(\tau)$, $i=1,\ldots 6$ are polynomials on $\tau$. Their
explicit form is not relevant for our purposes. The polynomials are
slightly different for each of the choices of $\kappa$, but of the
same order. Similarly, the components of the Weyl tensor at order
$p=3$ are of the form,
\begin{subequations}
\begin{eqnarray}
&& \phi_0^{(3)}=g_1(\tau)X_+X_+W_3 + g_2(\tau)mX_+X_+W_2 +g_3(\tau)(X_+W_1)^2 , \\
&& \phi_1^{(3)}=g_4(\tau)X_+W_3 + g_5(\tau)mX_+W_2 +
g_6(\tau)W_1X_+W_1 + g_7(\tau)m^2X_+W_1 , \\
&& \phi_2^{(3)}=g_8(\tau) W_3 + g_9(\tau) mW_2 + g_{10}(\tau) (W_1)^2
+ g_{11}(\tau)^2 m^2 W_1 + g_{12}(\tau) m^4 + g_{13}(\tau) b  ,\\
&& \phi_3^{(3)}=-g_4(-\tau)X_-W_3 - g_5(-\tau)mX_-W_2 -
g_6(-\tau)W_1X_-W_1 - g_7(-\tau)m^2X_-W_1 , \\
&& \phi_4^{(3)}=g_1(-\tau)X_-X_-W_3 + g_2(-\tau)mX_-X_-W_2 +g_3(-\tau)(X_-W_1)^2 . 
\end{eqnarray}
\end{subequations}
Again, the functions $g_i(\tau)$, $i=1,\ldots,13$ are polynomials,
while $b=2w_{1,2,0}w_{1,2,2}-w_{1,2,1}^2$ is a constant. 

The first new order corresponds to $p=4$. Here, again, the solutions
are still fully regular and polynomial:
\begin{subequations}
\begin{eqnarray}
&&
\phi_0^{(4)}=h_1(\tau)X_+W_1X_+W_2+h_2(\tau)W_1X_+X_+W_2+h_3(\tau)m(X_+W_1)^2
\nonumber \\
&&\phantom{XXXXXXXX}+h_4(\tau)mX_+X_+W_3+h_5(\tau)X_+X_+W_4  , \label{phi_order4a} \\
&& \phi_1^{(4)}= h_6(\tau)m^3X_+W_1+h_7(\tau)W_2X_+W_1
+h_8(\tau)W_1X_+W_2+h_9(\tau)mW_1X_+W_1 \nonumber \\
&&\phantom{XXXXXXXX}+h_{10}(\tau)mX_+W_3+h_{11}(\tau)X_+W_4, \\
&&
\phi_2^{(4)}=h_{12}(\tau)m^5+h_{13}(\tau)b+h_{14}(\tau)m^3W_1+h_{15}(\tau)W_1W_2+h_{16}(\tau)m(W_1)^2
 \nonumber \\
&&\phantom{XXXXXXXX}+h_{17}(\tau)mW_3+h_{18}(\tau)W_4 ,\\
&& \phi_3^{(4)}=-h_6(-\tau)m^3X_-W_1-h_7(-\tau)W_2X_-W_1
-h_8(-\tau)W_1X_-W_2-h_9(-\tau)mW_1X_-W_1 \nonumber \\
&&\phantom{XXXXXXXX}-h_{10}(-\tau)mX_-W_3-h_{11}(-\tau)X_-W_4 , \\
&&
\phi_4^{(4)}=h_1(-\tau)X_-W_1X_-W_2+h_2(-\tau)W_1X_-X_-W_2+h_3(-\tau)m(X_-W_1)^2
\nonumber \\
&&\phantom{XXXXXXXX}+h_4(-\tau)mX_-X_-W_3+h_5(-\tau)X_-X_-W_4. \label{phi_order4e}
\end{eqnarray}
\end{subequations}
Again, the functions $h_i(\tau)$, $i=1,\ldots,16$ are polynomials.

\subsection{The first obstructions to smoothness: orders $p=5,6$.}

The calculation of the solutions to the Bianchi transport equations up to
order $p=4$ have shown that all of them are polynomial, and thus
smooth at $I^\pm$. Consequently, the $v$ unknowns also happen to be
polynomial. The first modifications to this behaviour occur at the
rather high order $p=5$. Feeding the solution of the transport
equations up to order $p=4$ into the $v$ transport equations
(\ref{v_transport}) with $p=5$ and solving one finds again that the
components of $v^{(5)}$ are again polynomial. However,  the solution
of the Bianchi transport equations are of the form,
\begin{subequations}
\begin{eqnarray}
&& \phi_0^{(5)}=C_0 m^2 G^{(5)} \Bigl( (1-\tau)^7\ln(1-\tau) \nonumber
\\
&&\phantom{XXXXXX}+
  (1+\tau)^3(351-150\tau +48\tau^2-10\tau^3+\tau^4)\ln(1+\tau) \Bigr)
+ k_0(\tau), \label{P0_5} \\
&& \phi_1^{(5)}=C_1 m^2 G^{(5)} \Bigl( (1-\tau)^6(2\tau+5)\ln(1-\tau)
\nonumber \\
&&\phantom{XXXXXX}- (2\tau^3-15\tau^2+48\tau-75)(1+\tau)^4 \ln(1+\tau) \Bigr)+k_1(\tau) , \label{P1_5}\\
&& \phi_2^{(5)}=C_2 m^2 G^{(5)}\Bigl(
(1-\tau)^5(\tau^2+5\tau+8)\ln(1-\tau) \nonumber \\
&&\phantom{XXXXXX}+(1+\tau)^5 (\tau^2-5\tau+8)\ln(1+\tau) \Bigr) +
k_2(\tau), \label{P2_5} \\
&& \phi_3^{(5)}=C_1 m^2 G^{(5)} \Bigl(
  (1-\tau)^4(2\tau^3+15\tau^2+48\tau +75) \ln(1-\tau) \nonumber \\
&&\phantom{XXXXXX}-(1+\tau)^6(2\tau-5)\ln(1+\tau) \Bigr) + k_3(\tau) , \label{P3_5}\\
&& \phi_4^{(5)}=C_0 m^2 G^{(5)}  \Bigl(
  (1-\tau)^3(351+150\tau+48\tau^2+10\tau^3+\tau^4) \ln(1-\tau)
  \nonumber \\
&&\phantom{XXXXXX}+ (1+\tau)^7 \ln(1+\tau) \Bigr) + k_4(\tau), \label{P4_5}
\end{eqnarray}
\end{subequations}
where $C_0$, $C_1$, $C_3$ are non-relevant non-zero numerical factors,
$k_i(\tau)$ $i=0,\ldots,4$ are polynomials analogous to, for example, those in
(\ref{phi_order4a})-(\ref{phi_order4e}), depending on $m$, $W_1$,
$W_2$, $W_3$, $W_4$, their $X_\pm$ derivatives and products of
them. Most remarkably,
\begin{equation}
G^{(5)}=\sum_{k=0}^4 G^{(5)}_k T_{4\phantom{k}2}^{\phantom{4}k},
\end{equation}
where,
\begin{subequations}
\begin{eqnarray}
&& G^{(5)}_0= m w_{2,4,0}-2\sqrt{6}w_{1,2,0}^2, \label{G50} \\
&& G^{(5)}_1= m w_{2,4,1}-4\sqrt{3}w_{1,2,0}w_{1,2,1}, \label{G51} \\
&& G^{(5)}_2= m w_{2,4,2}-4w_{1,2,1}^2-4w_{1,2,0}w_{1,2,2}, \label{G52}
\\
&& G^{(5)}_3= m w_{2,4,3}-4\sqrt{3}w_{1,2,1}w_{1,2,2}, \label{G53} \\
&& G^{(5)}_4= m w_{2,4,4}-2\sqrt{6}w_{1,2,2}^2 \label{G54}.
\end{eqnarray}
\end{subequations}
Thus, the coefficients $G^{(5)}_k$ are (besides an irrelevant
numerical factor) the Newman-Penrose constants of the development of
the time symmetric conformally flat initial data ---\cite{FriKan00}. 

Plugging the solutions (\ref{P0_5})-(\ref{P4_5}) into the transport
equations for $p=6$ one finds that the solutions of the sectors of the
form $T_{4\phantom{k}j}^{\phantom{4}k}$ develop terms containing
$\ln^2(1-\tau)$ and $\ln(1-\tau)$. Furthermore, the sectors
$T_{6\phantom{k}j+1}^{\phantom{6}k}$ in $\phi^{(p)}_j$ contain
logarithms. More precisely, the solution will be of the form,
\begin{eqnarray}
&&\phi_j^{(6)}=G^{(5)} \Bigl( l_{1_j}(\tau)\ln^2(1-\tau) + l_{2_j}(\tau)
\ln(1-\tau) + l_{3_j}(\tau)\ln^2(1+\tau) + l_{4_j}(\tau)
\ln(1+\tau) \Bigr) \nonumber \\
&&\phantom{XXXXX}+ G^{(6)}\Bigl( l_{5_j}(\tau)\ln(1-\tau) +
l_{6_j}(\tau)\ln(1+\tau) \Bigr) + l_{7_j}(\tau) , 
\end{eqnarray}
where $j=0,\ldots,7$, and $l_{k_j}(\tau)$ are polynomials. The
coefficients $G^{(6)}$ are new obstructions to the smoothness of the
solutions. These will be discussed a bit later. It is not hard to
imagine that from this point onwards, terms containing $\ln(1-\tau)$,
$\ln(1+\tau)$ and higher order powers of them will spread all around the
solutions to the transport equations. Instead of analysing this
phenomenon, we will rather focus on the smooth solutions. 

Setting the Newman-Penrose constants to zero, the solution at order
$p=6$ is of the form, 
\begin{subequations}
\begin{eqnarray}
&& \phi_0^{(6)}=D_0 m^3 G^{(6)} \Bigl(
(2+\tau)(1-\tau)^8\ln(1-\tau) \nonumber \\
&&\phantom{XXXX}+(-254+233\tau-128\tau^2+46\tau^3-10\tau^4+\tau^5)(1+\tau)^4\ln(1+\tau)
\Bigr) + l_0(\tau)   , \\
&& \phi_1^{(6)}=D_1 m^3 G^{(6)} \Bigl(
(23+20\tau+5\tau^2)(1-\tau)^7\ln(1-\tau)
\nonumber \\
&&\phantom{XXXX}-(233-256\tau+138\tau^2-40\tau^3+5\tau^4)(1+\tau)^5\ln(1+\tau)\Bigr)+l_1(\tau) , \\
&& \phi_2^{(6)}=D_2 m^3 G^{(6)}\Bigl(
(64+69\tau+30\tau^2+5\tau^3)(1-\tau)^6\ln(1-\tau) \nonumber \\
&&\phantom{XXXX}+(-64+69\tau)-30\tau^2-5\tau^3)(1+\tau)^6\ln(1+\tau)\Bigr)
+ l_2(\tau), \\
&& \phi_3^{(6)}=D_1 m^3 G^{(6)} \Bigl(
(233+256\tau+138\tau^2+40\tau^3+5\tau^4)(1-\tau)^5\ln(1-\tau)
\nonumber \\
&&\phantom{XXXX}+(23-20\tau+5\tau^2)(1+\tau)^7\ln(1+\tau)
\Bigr) + l_3(\tau) , \\
&& \phi_4^{(6)}=D_0 m^3 G^{(6)}  \Bigl(
(254+233\tau+128\tau^2+46\tau^3+10\tau^4+\tau^5)(1-\tau)^4\ln(1-\tau)
\nonumber \\
&&\phantom{XXXX}+(-2+\tau)(1+\tau)^8\ln(1+\tau)\Bigr) + l_4(\tau),
\end{eqnarray}
\end{subequations}
with $l_0(\tau),\dots,l_4(\tau)$ polynomials, and 
\begin{equation}
G^{(6)}= \sum_{k=0}^6 G^{(6)}_k T_{6\phantom{k}3}^{\phantom{6}k},
\end{equation}
where the coefficients $G^{(6)}_k$, $k=0,\dots,6$ are given in terms
of initial data quantities by,
\begin{subequations}
\begin{eqnarray}
&& G^{(6)}_0=m^2 w_{3,6,0}-12\sqrt{10}w_{1,2,0}^3, \label{G60} \\
&& G^{(6)}_1=m^2 w_{3,6,1}-12\sqrt{30}w_{1,2,0}^2w_{1,2,1},  \label{G61}\\
&& G^{(6)}_2=m^2
w_{3,6,2}-24\sqrt{6}w_{1,2,0}w_{1,2,1}^2-12\sqrt{6}w_{1,2,0}^2w_{1,2,2},
\label{G62} \\
&& G^{(6)}_3=m^2 w_{3,6,3} -72 w_{1,2,0} w_{1,2,1} w_{1,2,2} -24
w_{1,2,1}^3,  \label{G63}\\
&& G^{(6)}_4=m^2 w_{3,6,4} -24\sqrt{6} w_{1,2,1}^2 w_{1,2,2}
-12\sqrt{6}w_{1,2,0} w_{1,2,2}^2, \label{G64}\\
&& G^{(6)}_5=m^2 w_{3,6,5} -12\sqrt{30} w_{1,2,1}w_{1,2,2}^2,
\label{G^5}\\
&& G^{(6)}_6=m^2 w_{3,6,6} -12\sqrt{10} w_{1,2,2}^3. \label{G66}
\end{eqnarray}
\end{subequations}
In analogy to the order $p=5$ we will refer to these coefficients as
to the \emph{order $6$ Newman-Penrose ``constants''}. One is naturally
bound to ask whether the coefficients $G^{(6)}$ are actually
associated to some conserved quantities at null infinity in the same
way that the coefficients $G^{(5)}$ are. This consideration is beyond
the scope of the present article, and will be analysed in detail in
future work. It is pointed out, as a plausibility
argument, that in the analysis of polyhomogeneous Bondi expansions
carried out in for example \cite{ChrMacSin95,Val98,Val99a}, the first
logarithmic terms appearing in the expansions were associated with a
conserved quantity on null infinity. As it can be seen from our
discussion, if $G^{(5)}=0$, then the first logarithmic terms appearing
in our expansions are precisely those associated to $G^{(6)}$.
Notwithstanding, the expansions described here are based in the
conformal geodesics gauge, while those in
\cite{ChrMacSin95,Val98,Val99a} use the so-called Bondi gauge. Thus,
one would have to look in detail the possible appearance of
logarithmic terms in the transformation connecting the two gauges.

 \begin{main}[Main theorem, precise formulation]
 Necessary conditions for the development of initial data which is time
 symmetric and conformally flat in a neighbourhood $B_a(i)$ of (spatial)
 infinity to be smooth on the set $I\cup I^+ \cup I^-$ are that
 the Newman-Penrose constants, $G^{(5)}_k$, $k=0,\ldots,5$, should
 vanish. Furthermore, the ``higher order'' Newman-Penrose constants,
 $G^{(6)}_k$, $k=0,\ldots,6$, should also vanish. If only the
 coefficients $G^{(5)}_k$ vanish then the rescaled Weyl spinor is at most
 $C^2$ on a neighbourhood of either $I^+$ or $I^-$. If both $G^{(5)}$
 and $G^{(6)}$ vanish then generically the Weyl tensor will be at most
 of class $C^3$ of the latter neighbourhoods.
 \end{main}

From the last result one can extract directly the following (important)
consequence:

\begin{corollary}
The regularity condition (\ref{regcond}) is not a sufficient
condition for the smoothness at $I\cup I^+ \cup I^-$ of the development of
asymptotically Euclidean, time symmetric initial data sets.
\end{corollary}

This corollary is, thus, a negative answer to the conjecture raised in
\cite{Fri02a}. It is nevertheless surprising that the obstructions to
the smoothness of null infinity arise at such a high order in the
expansions. In order to acquire a deeper understanding of why this is
the case would require in turn an abstract understanding of the
algebraic structure of the transport equations (\ref{v_transport}) and
(\ref{b_transport}). It is not unreasonable to reckon that some group
theoretical properties of the whole set up play a major role here. 

The theorem also suggest the following un expected conjecture:

\begin{conjecture}
The developments of the Brill-Lindquist and Misner initial data sets
possess non-smooth null infinities. 
\end{conjecture}

In \cite{DaiVal02,Val02b} the Newman-Penrose constants of the
Brill-Lindquist and Misner data sets \cite{BriLin63,Mis63} have been
calculated using the formula found by Friedrich \& K\'{a}nn\'{a}r
\cite{FriKan00}. Due to the axial symmetry, there is only one
non-vanishing Newman-Penrose constant. Furthermore, the constant
vanishes if and only if the data sets are actually the Schwarzschild
initial data. That the development of these two initial data sets
possess a non-smooth null infinity may have implications in the
description of their late time behaviour in terms of linear
perturbations on a Schwarzschild background.

Some remarks regarding the theorem come also into place:

\textbf{Remark 1.} The calculations for the order $p=7$ are already
beyond the capabilities of {\tt Maple V} ---the expressions involved
are too large for the simplification routines of the computer algebra
system, even for an Origin computer available at the Albert Einstein
Institute. Nevertheless, the calculations of axially symmetric
situations are still possible. These have been carried out for the
orders $p=7$ and $p=8$ inclusive. Assuming that both $G^{(5)}$ and
$G^{(6)}$ vanish, the solutions of the Bianchi
transport equations have again the expected form:
\begin{equation}
\phi_j^{(7)}=E_jm^3 G^{(7)} \Bigl( m_{1_j}(\tau)\ln(1-\tau) +
m_{2_j}(\tau) \Bigr) +m_{3_j}(\tau),
\end{equation}
where now due to the axial symmetry there is only one order $7$
Newman-Penrose constant,
\begin{equation}
G^{(7)}= \Bigl( m^3 w_{4,8,4} -192 w_{1,2,1}^4 \Bigl)T_{8\phantom{4}4}^{\phantom{8}4}.
\end{equation}
If in turn $G^{(7)}$ vanishes,
\begin{equation}
\phi_j^{(8)}=F_jm^4 G^{(8)} \Bigl( n_{1_j}(\tau)\ln(1-\tau) +
n_{2_j}(\tau) \Bigr) +n_{3_j}(\tau),
\end{equation}
with,
\begin{equation}
G^{(8)}= \Bigl( m^4 w_{5,10,5} -1920 w_{1,2,1}^5 \Bigl)T_{10\phantom{5}5}^{\phantom{10}5}.
\end{equation}
From this evidence it is not too hard to guess the
following  general formula for the obstructions of the smoothness of
null infinity in the axial situation,
\begin{equation}
\label{axial_higher_np}
G^{(p)}= \Bigl( m^{p-4} w_{p-3,2p-6,p-3} - 2^{p-4}(p-4)!
(w_{1,2,1})^{p-3} \Bigl) T_{2p-6\phantom{p-3}p-3}^{\phantom{2p-6}p-3}.
\end{equation}
The proof of such a formula is nevertheless beyond our current
understanding of the transport equations. The significance of the
latter expression will be discussed in short.

\textbf{Remark 2.} In the non-axially symmetric case, it is
conjectured that obstructions to the smoothness of null infinity
(Generalised Newman-Penrose constants) are given in terms of the
initial data by the following expressions,
\begin{equation}
\label{nonaxial_higher_np}
G^{(p)}_k=m^{p-4} w_{p-3,2p-6,k}-\sum_{s_0+s_1+s_2=k} c_{p,k;s_0,s_1,s_2}w_{1,2,0}^{s_0}w_{1,2,1}^{s_1}w_{1,2,2}^{s_2},
\end{equation}
with $k=0,1,\dots,p$, where the coefficients $c_{p,k;s_0,s_1,s_2}$ are
some numerical constants.

\subsection{Obstructions to smoothness and the Schwarzschild initial
  data.}

As mentioned before, the Newman-Penrose constants of the Schwarzschild
spacetime are all zero. Thus, in order to gain some insight into the
significance of the expressions (\ref{axial_higher_np}) and
(\ref{nonaxial_higher_np}) it is convenient to see what occurs in the case of the Schwarzschild initial data.

It is not complicate ---although certainly messy--- to calculate an
expression for initial data for the Schwarzschild on the slice of time
symmetry. On this slice the initial 3-metric is conformally
flat. Thus, the required (harmonic) conformal factor can be calculated
directly from the Green
function for the three dimensional Laplace equation in spherical
coordinates. The Green function is given by,
\begin{equation}
G(r,\theta,\phi;r',\theta',\phi')=\sum_{n=0}^\infty\sum_{m=-n}^n\frac{4\pi}{2n+1}\overline{Y_{n,m}(\theta,\phi)}Y_{n,m}(\theta',\phi')\frac{r'^n}{r^{n+1}},
\end{equation}
for $r>r'$.

The latter expression can be lifted into the frame bundle
$C_{a,\kappa}$ and written in terms of the functions
$T_{j\phantom{k}l}^{\phantom{j}k}$ using formula (\ref{Y_to_T}) to get,
\begin{equation}
W=\frac{m}{2}\sum_{n=0}^\infty\sum_{k=0}^{2n} r'^n
T_{2n\phantom{2n-k}n}^{\phantom{2n}2n-k}(t') \;
T_{2n\phantom{k}n}^{\phantom{2n}k} \rho^n.
\end{equation}
where $(t',r')$ denote the coordinates of the singularity of the Green
function on the frame bundle. Thus, the
$T_{2n\phantom{2n-k}n}^{\phantom{2n}2n-k}(t')$ are fixed complex
numbers. We write,
\begin{equation}
w_{n,2n,k}=\frac{1}{2} m r'^n n!
T_{2n\phantom{2n-k}n}^{\phantom{2n}2n-k}(t'),
\end{equation}
for $n=0,1,\ldots$ and $k=0,\ldots,2n$. The latter coefficients are
not independent but related to each other via recurrence relations,
like
\begin{eqnarray}
&&(n+1)(2n+1) w_{1,2,1}\times w_{n,2n,k}
=\sqrt{(k+1)(2n-k+1)}w_{n+1,2(n+1),k+1} \nonumber \\
&&\phantom{XXXXXXX}+r'^2 n(n+1)\sqrt{k(2n-k)} w_{n-1,2(n-1),k-1}.
\label{recurrence}
\end{eqnarray}
These can be readily obtained from similar recurrence relations
holding for the spherical harmonics. The Schwarzschild solution has an
obvious axial symmetry. If one considers an orientation of the
coordinate system in a way that makes this symmetry of the initial data explicit ---the singularity of the Green function is set along the $z$
axis--- one ends up with a much simplified expression,
\begin{equation}
W=\sum_{n=0}^\infty \left(\frac{2}{m}\right)^{n-1}(w_{1,2,1})^n
T_{2n\phantom{n}n}^{\phantom{2n}n}\rho^n,
\end{equation}
so that the only non-vanishing $w_{n,2n,k}$ coefficients are given by,
\begin{equation}
w_{n,2n,n}=\left(\frac{2}{m}\right)^{n-1}(w_{1,2,1})^n n!\;.
\end{equation}
One obtains precisely this expression if one requires that the
hierarchy of axial Newman-Penrose constants, formula (\ref{axial_higher_np}),
vanishes at every order $p$. It is however important to point out that
this is not a proof but again a plausibility argument as formula
(\ref{axial_higher_np}) has only been verified up to order $p=8$. A
much more involved calculation shows that the expressions one obtains
for the coefficients $w_{2,4,0},\ldots,w_{2,4,4}$ and
$w_{3,6,0},\ldots,w_{3,6,6}$ by setting $G^{(5)}$ and $G^{(6)}$ to
zero are precisely those one has for the Schwarzschild initial
data. This involves the use of the recurrence relation
(\ref{recurrence}). 

Now, the fuction $W$ is a solution to $\Delta W=0$, and thus
analytic. This in turn implies that the function $W$ one deduces from
requiring the coefficients (\ref{axial_higher_np}) to vanish at all
orders $p$ is exactly Schwarzschildean. This evidence leads to the following, 

\begin{conjecture} [Precise formulation]
  For every $k>0$ there exists a $p=p(k)$ such that the time evolution
  of an asymptotically Euclidean, time symmetric, conformal flat,
  conformally smooth initial data set admits a conformal extension to
  null infinity of class $C^k$ near spacelike infinity, if and only if
  the initial data set is Schwarzschildean to order $p(k)$.
\end{conjecture}

The latter is the successor of the conjecture, in the context of
time symmetric conformally flat, of the conjecture by Friedrich
\cite{Fri02a} to which reference was made in the introduction.

\section{Conclusions and extensions}

The results presented in the main theorem together with the
considerations leading to the conjecture put forward in section 5 seem
to suggest that no gravitational radiation should be present around
spatial infinity if one is to have a smooth null infinity. In other
words, the notion of smooth null infinity seems to be incompatible
with the presence of radiation around $i^0$. Whether this behaviour of
spacetime in the neighbourhood of spatial infinity has some
implications on either the demeanour of the sources of the gravitational
field in the infinite past or on the nature of incoming radiation
travelling from past null infinity is a naturally arising question.   

A natural extension of the calculations here described would be to
consider what happens with more general (i.e. not conformally flat)
time symmetric initial data. If the time symmetric initial data
is not conformally flat, then the Newman-Penrose constants are given
in terms of the initial data by,
\begin{eqnarray*}
&&\widetilde{G}^{(5)}_0= m w_{2,4,0}-2\sqrt{6}w_{1,2,0}^2-\frac{1}{254\sqrt{6}}R^{(2)}_0, \nonumber \\
&&\widetilde{G}^{(5)}_1= m w_{2,4,1}-4\sqrt{3}w_{1,2,0}w_{1,2,1}-\frac{1}{254\sqrt{6}}R_1^{(2)}, \nonumber \\
&&\widetilde{G}^{(5)}_2= m w_{2,4,2}-4w_{1,2,1}^2-4w_{1,2,0}w_{1,2,2}-\frac{1}{254\sqrt{6}}R_2^{(2)}, \nonumber
\\
&&\widetilde{G}^{(5)}_3= m w_{2,4,3}-4\sqrt{3}w_{1,2,1}w_{1,2,2}-\frac{1}{254\sqrt{6}}R_3^{(2)}, \nonumber \\
&&\widetilde{G}^{(5)}_4= m w_{2,4,4}-2\sqrt{6}w_{1,2,2}^2-\frac{1}{254\sqrt{6}}R_4^{(2)},
\end{eqnarray*}
where $R_k^{(2)}$ are coefficients appearing in the expansion of the
Ricci scalar $r$ of the initial $3$-metric, around spatial infinity.
In the view of the present results one would expect to have the
quantities $\widetilde{G}^{(5)}_k$ appearing as obstructions to the
smoothness of null infinity.  Similarly, one would expect to obtain
generalisations $\widetilde{G}^{(6)}_k$ of the higher order
Newman-Penrose constants $G^{(6)}_k$ by adding suitable expressions
containing the Ricci scalar. Ultimately, one would like to prove that
there is an infinite hierarchy of such quantities as obstructions to
the smoothness of $\scri$. The relation of these quantities with
static initial data is to be analysed, at least for the first orders.
It could well be the case that a time symmetric initial data set will
yield a development with smooth null infinity only if it is static in
a neighbourhood of spatial infinity. In relation to this, it is noted
that the asymptotically Euclidean static initial data satisfy the
regularity condition (\ref{regcond}) ---see \cite{Fri88}.

With regard to a proof to the conjecture presented in this article, it
is noted that it would require a much deeper understanding of the
properties of the transport equations at spatial infinity that the one
currently available. In particular one would like to be able to
discuss its algebraic properties in a much more abstract way, and with
out having to resort to ``{e}xplicit expressions'' as it was done
here. As mentioned before, group
theoretical properties of the set up should play a crucial role here. 

\subsection*{Acknowledgements} 
I thank H. Friedrich who suggested the problem and helped with
discussions and clarifications all along the long winding road. I also
thank S. Dain and M. Mars from which I have benefited
through several discussions.

\section*{Appendix: on the Computer Algebra implementation}

The calculations described in this article are hefty, however quite
amenable to the use of computer algebra systems. In the present case the
computer algebra system {\tt Maple V} was used. The main source of
difficulties in solving the transport equations implied by the conformal field
equation on the cylinder at spatial infinity is not the actual procedure of
solving the equations, but producing the equations that have to be
solved. To understand why this is the case, we recall the quasilinear nature
of the transport equations. Accordingly, in the construction of the equations
for the diverse spherical sectors of the various unknowns one encounters
products of the functions $T_{j\phantom{k}l}^{\phantom{j}k}$, which have to be
expanded using the formula (\ref{monster}), which in turn requires the
computation of several Clebsch-Gordan coefficients. The number of these
products heavily increases as one considers higher and higher expansions
orders. There are {\tt Maple V} scripts available in the web. We have made use
of those written by A. B. Coates \cite{Coa}.

For the purposes of the computer algebra implementation, the diverse
spinor arising have been considered as arrays running on the set of
indices $\{0,1\}$. In this way all the elementary spinors introduced
in section 2.1 can be readily defined. For example, the spinor
$x_{ab}$ would be encoded as,
\[
{\tt >x[0,0]:=0: x[0,1]:=1/sqrt(2): x[1,0]:=1/sqrt(2): x[1,1]:=0:}
\]
The functions $T_{j\phantom{k}l}^{\phantom{j}k}$ have also been
considered as arrays, {\tt T[j,k,l]} with empty entries, so that
they can be manipulated formally as symbols (names, in the {\tt Maple
  V} terminology). With this idea in mind, scripts that evaluate
$X_\pm$ derivatives according to the usual rules and take complex
conjugates of expressions containing
$T_{j\phantom{k}l}^{\phantom{j}k}$ have been written. Two further
scripts, one implementing the formula (\ref{monster}) to express
products of $T_{j\phantom{k}l}^{\phantom{j}k}$'s as linear
combinations of $T_{j\phantom{k}l}^{\phantom{j}k}$'s, {\tt Texpand},
and another that given a expression collects the terms possessing the
same $T_{j\phantom{k}l}^{\phantom{j}k}$, {\tt Tcollect}, are the core
of our implementation.

In order to compute the expansions up to a given order $p$ one
proceeds in two steps. Firstly, one has to compute the initial data
for the conformal propagation equations. As described in the main text
we have restricted our implementation to time symmetric, conformally
flat initial data. The computer algebra construction of time symmetric
data satisfying the regularity condition (\ref{regcond}) is a project
on its own right that we expect to address in the near future. In the
case of time symmetric, conformally flat data the whole free data is
contained in a function $W$ solution of the Laplace equation. In order
to calculate the solution of the transport equations up to say, order
$p$ one needs to expand the function $W$ in $\rho$ up to order $p$.
Thus, the main input of the script that calculates the initial data
for the transport equations are the essentially the order of expansion
{\tt p} and the function {\tt kappa} corresponding to the remaining
conformal freedom available in our setting. Two choices have been
used, $\kappa=\rho$ and $\kappa=\omega$. The function $W$ is provided
as an object of type {\tt series} in the form,
\begin{eqnarray*}
&&{\tt
 >W:=m/2*T[0,0,0]+sum('sum('(1/(i)!)*w[i,2*i,k]*T[2*i,k,i]*rho\hat{\phantom{x} }i',} \\
&&{\tt 'k'=0..2*i)','i'=1..6)+O(rho\hat{\phantom{x}}7):}
\end{eqnarray*}
Two versions of the script have been written: one which deals with axially
symmetric initial data, and the other dealing with general time symmetric,
conformally flat initial data, i.e. data which assumes no symmetries. The
computations carried in these scripts are essentially completely automatised
(i.e. no human intervention is required), and their output is saved directly.

The actual computation and solving of the transport equations is carried out
in a second stage. Given an order {\tt p} the scripts load first the
results of the previous orders, and then proceeds to the most
computationally expansive part of the script: the calculation of the
transport equations at order {\tt p}. As discussed in the main text,
for a given order one needs first to calculate and solve the so-called
$v$ equations. Using our definitions of spinors as arrays, the
transport equations can be readily coded. For example, equation
(\ref{p1}) would be coded as:
\begin{eqnarray*}
&&{\tt >for\;\;a\;\;in\;\; \{0,1\}\;\; do} \\
&&{\tt >\phantom{fora}for\;\;b\;\;in\;\;\{0,1\}\;\;do} \\    
&&{\tt >\phantom{foraforb}
  eqn1[p,a,b]:=-diff(c0[p,a,b],tau)} \\
&&{\tt
  \phantom{>foraforb}-sum('binomial(p,k)*sum('sum('sum('sum('eup[e,r]*eup[s,t]*chi[k,a,b,r,t]} \\
&&{\tt \phantom{>foraforb}*c0[p-k,e,s]','e'=0..1)','r'=0..1)','s'=0..1)','t'=0..1)','k'=0..p)-f[p,a,b]:} \\
&&{\tt >\phantom{fora}od:} \\
&&{\tt >od:} 
\end{eqnarray*}
The remaining equations, (\ref{p2}) to (\ref{p7}), have been encoded
in a similar way. The procedure {\tt Tcollect} is then used to write
all the equations in the form of linear combination of the {\tt
  T[j,k,l]}'s. Using, {\tt Tcollect} the resulting expressions are
then grouped. A further script then collects all the equations
corresponding to a given mode and picks up the corresponding initial
data. The systems of equations which in the generic case consists of
45 linear ordinary differential equations is using the {\tt dsolve}
procedure of {\tt Maple V} with the {\tt series} option set on. In this way
{Maple V} calculates the solutions directly as series expansions. The
order of the expansion is set sufficiently high, {\tt order:=60}, so
that the series can truncate if the solutions correspond to
polynomials. Once the solutions have been calculated, the solutions
are saved and the memory of the system is cleared. 

The transport equations arising from the Bianchi identities,
(\ref{b0})-(\ref{b4}), are likewise encoded and solved. As a check of
correctness, the constraint equations (\ref{c1})-(\ref{c3}), have also
been programmed. After the solutions to the
equations (\ref{b0})-(\ref{b4}) have been obtained, these are
substituted in the equations (\ref{c1})-(\ref{c3}) which should be
satisfied automatically. As discussed in the main text, at order $p=5$
and higher the solutions to some of the spherical sectors contain
logarithms. These are firstly detected in the {\tt Maple V}
computations as series solutions that do not truncate. For the cases
when this occurs, some supplementary scripts have been written. These
essentially implement the analysis carried in the section 4.1. The
scripts calculates the fundamental matrix of the system, and uses it
to calculate the solution in a closed form so that the logarithms
present in the expression are explicit.

In order to give an impression of the progressive calculational
complexity of the whole procedure, it is noted that the calculations
in the axially symmetric case for the orders $p=1,\ldots,8$ have taken
$9.6$, $36$, $104.2$, $284.4$, $894$, $2917.9$, $9605$, and $32139$
seconds respectively in a computer with a Dell Xeon computer (processor
speed of about $2 GHz$ and $2Gb$ of RAM memory). The files (text only)
containing the order $p=6$ v-equations and b-equations are $2.2Mb$ and
$0.7Mb$ big respectively. 


\end{document}